\documentclass[12pt]{article}
\usepackage{epsfig,cite}
\usepackage{axodraw}
\include{epsf}
\newcount\timecount
\newcount\hours \newcount\minutes  \newcount\temp \newcount\pmhours
\hours = \time
\divide\hours by 60
\temp = \hours
\multiply\temp by 60
\minutes = \time
\advance\minutes by -\temp
\def\hour{\the\hours}
\def\minute{\ifnum\minutes<10 0\the\minutes
            \else\the\minutes\fi}
\def\clock{
\ifnum\hours=0 12:\minute\ AM
\else\ifnum\hours<12 \hour:\minute\ AM
      \else\ifnum\hours=12 12:\minute\ PM
            \else\ifnum\hours>12
                 \pmhours=\hours
                 \advance\pmhours by -12
                 \the\pmhours:\minute\ PM
                 \fi
            \fi
      \fi
\fi
}

\def\monthname{\relax\ifcase\month 0/\or January\or February\or
   March\or April\or May\or June\or July\or August\or September\or
   October\or November\or December\else\number\month/\fi}

\def\bold#1{\setbox0=\hbox{$#1$}%
     \kern-.025em\copy0\kern-\wd0
     \kern.05em\copy0\kern-\wd0
     \kern-.025em\raise.0433em\box0 }

\def\beq{\begin{equation}}
\def\eeq{\end{equation}}
\def\bear{\begin{eqnarray}}
\def\eear{\end{eqnarray}}

\def\ss{\scriptscriptstyle}
\def\ga{\mathrel{\raise.3ex\hbox{$>$\kern-.75em\lower1ex\hbox{$\sim$}}}}
\def\la{\mathrel{\raise.3ex\hbox{$<$\kern-.75em\lower1ex\hbox{$\sim$}}}}
\def\gev{{\rm \, Ge\kern-0.125em V}}
\def\tev{{\rm \, Te\kern-0.125em V}}
\def\gyr{{\rm \, G\kern-0.125em yr}}

\def\half{{\textstyle{1\over2}}}

\def\slash#1{\rlap{\hbox{$\mskip 1 mu /$}}#1}%
\def\nl{\hfill\nonumber\\&&}
\def\nnl{\hfill\nonumber\\}

\def\ttbt{\tan^2 \beta}

\def\gappeq{\mathrel{\rlap {\raise.5ex\hbox{$>$}}
{\lower.5ex\hbox{$\sim$}}}}
\def\lappeq{\mathrel{\rlap{\raise.5ex\hbox{$<$}}
{\lower.5ex\hbox{$\sim$}}}}
\def\Toprel#1\over#2{\mathrel{\mathop{#2}\limits^{#1}}}

\def\stau{\widetilde \tau}

\def\snu{\widetilde \nu}

\def\bal{\begin{array}}
\def\eal{\end{array}}

\def\beqn{\begin{eqnarray}}
\def\eeqn{\end{eqnarray}}

\def\mchi{m_{\tilde \chi}}
\def\msnu{m_{\tilde\nu}}
\def\m12{m_{1\!/2}}

\def\mz{m_{\ss Z}}

\def\mstau{m_{\tilde{\ell}_1}}

\def\msnu{m_{\tilde{\nu}}}

\def\PR{{Phys.~Rev.} }

\def\stau{\tilde{\tau}}
\def\mstau{m_{\tilde{\tau}}}
\def\mgrav{m_{\widetilde G}}

\def\mchi{m_{\chi}}

\def\bea{\begin{eqnarray}}
\def\eea{\end{eqnarray}}

\begin{document}
\begin{titlepage}
\pagestyle{empty}
\baselineskip=21pt
\renewcommand{\thefootnote}{\fnsymbol{footnote}}
\rightline{CERN-PH-TH/2008-159}
\rightline{UMN--TH--2708/08}
\rightline{FTPI--MINN--08/28}
\rightline{IPPP/08/53}
\rightline{DCPT/08/106}
\vskip 0.2in
\begin{center}
{\large{\bf Sneutrino NLSP Scenarios in the NUHM with Gravitino Dark Matter}}
\end{center}
\begin{center}
\vskip 0.2in
{\bf John~Ellis}$^1$, {\bf Keith~A.~Olive}$^{2}$
and {\bf Yudi Santoso}$^{3}$
\vskip 0.1in

{\it
$^1${TH Division, CERN, Geneva, Switzerland}\\
$^2${William I. Fine Theoretical Physics Institute, \\
University of Minnesota, Minneapolis, MN 55455, USA}\\
$^3${Institute for Particle Physics Phenomenology, \\
Durham University, Durham DH1 3LE, UK}}

\vskip 0.2in
{\bf Abstract}
\end{center}
\baselineskip=18pt \noindent
We analyze scenarios in which some flavour of sneutrino is the next-to-lightest
supersymmetric particle (NLSP), assuming that the gravitino is the lightest
supersymmetric particle (LSP) and provides the cold dark matter. 
Such scenarios do not arise in the constrained
supersymmetric extension of the Standard Model (CMSSM) with universal
gaugino and scalar masses input at the GUT scale. However, models with
non-universal Higgs masses (NUHM) do allow scenarios with a sneutrino NLSP,
which are quite generic. We illustrate how such scenarios may arise, analyze the
possible metastable sneutrino lifetime, and explore the theoretical,
phenomenological, experimental and cosmological constraints on such scenarios.
We also discuss the collider signatures of such scenarios, how they may be
distinguished from neutralino LSP scenarios, and how different flavours of
sneutrino NLSP may be distinguished.

\bigskip
\leftline{CERN-PH-TH/2008-159}
\leftline{July 2008}
\bigskip
\bigskip
\end{titlepage}
\baselineskip=18pt
\renewcommand{\thefootnote}{\arabic{footnote}}

\section{Introduction}

In the framework of supersymmetry with conserved $R$ parity, the lightest
supersymmetric particle (LSP) can have neither electromagnetic nor strong
interactions: otherwise it would have bound to conventional matter and been
detected in searches for anomalous heavy nuclei \cite{isotopes}. 
Within the minimal supersymmetric extension
of the Standard Model (MSSM), weakly-interacting candidates for the LSP are the
lightest sneutrino $\tilde\nu$, the lightest neutralino $\chi$, and the
gravitino ${\tilde G}$. The (left-`handed') sneutrino LSP hypothesis is excluded
by a combination of neutrino counting at LEP and direct dark matter searches \cite{FOS94}.
Accordingly, general attention is focused on the neutralino (NDM) \cite{EHNOS} and gravitino
dark matter (GDM) \cite{gdm,FengGDM} possibilities, and in this paper we assume the latter.

The next question is the possible nature of the next-to-lightest supersymmetric
particle (NLSP) in a GDM model. Two natural possibilities are the other candidates 
for the LSP, namely the sneutrino and the neutralino, but
the NLSP could equally well be charged and even coloured. Indeed, 
the lighter stau slepton is a natural candidate for the NLSP~\cite{FengGDM3}
within the constrained MSSM (CMSSM) with gravity-mediated supersymmetry
breaking, in which the soft supersymmetry-breaking scalar masses $m_0$,
trilinear parameters $A_0$ and gaugino masses $m_{1/2}$ are each assumed to be
universal at a GUT input scale \cite{eoss,cmssmwmap}.  
The lighter stau is also a natural possibility within minimal supergravity
(mSUGRA), in which the gravitino mass is fixed: $m_{\tilde G} = m_0$, and there
is an additional relation between trilinear and bilinear soft
supersymmetry-breaking parameters \cite{vcmssm}. Another possibility for the NLSP
within a scenario with non-universal Higgs masses (NUHM) \cite{nonu,ourNUHM1,ourNUHM2,baerNUHM,eosk4} is the lighter stop~\cite{deos}.

In this paper we study the other possibility for the NLSP within the NUHM
with gravity-mediated supersymmetry breaking, namely
the lightest sneutrino, assuming that the gravitino provides the cold dark 
matter~\footnote{The possibility of a sneutrino NLSP has also
been studied within gaugino-mediated models of supersymmetry 
breaking~\cite{BCKS}.}. Like other NLSP candidates in gravity-mediated
scenarios, the sneutrino NLSP within the NUHM is expected to be very long-lived.
The dominant decays of the other NUHM NLSP candidates produce particles with
copious interactions such as charged particles and photons, which are subject to
strong cosmological limits \cite{decay1}. 
These limits are 
strong enough to exclude effectively all of the parameter space
where the lightest neutralino is the NLSP \cite{CEFOS}~\footnote{If
the gravitino mass is much lighter than the neutralino mass, these limits might
still be satisfied.}.
However, the dominant decay mode of the sneutrino is
${\tilde \nu} \to {\tilde G} \nu$, and the cosmological limits on neutrino
injection are much weaker than those on the injection of photons and charged
particles 
\cite{KKKM-1,KKKM-2}. Therefore the cosmological limits on NUHM ${\tilde \nu}$ NLSP
scenarios are relatively weak, leaving considerable scope for a sneutrino
NLSP. On the other hand, the sneutrino must appear
at the end of the decay chain of every MSSM sparticle produced at a collider,
and the particles produced in supersymmetric decay cascades provide
distinctive experimental signatures for a sneutrino NLSP \cite{CK}. In particular, the charged
leptons produced in association with a ${\tilde \nu}$ NLSP provide tools
for diagnosing its flavour.

The structure of this paper is as follows. In Section~2 we discuss sneutrino
properties within the NUHM, including its mass and lifetime. In Section~3 we
discuss the relic abundance of sneutrinos after freeze-out from a primordial
plasma in thermal equilibrium. In Section~4 we analyze the NUHM parameter
space and identify regions where the NLSP may be either the ${\tilde \nu_{e,\mu}}$
or the ${\tilde \nu_\tau}$. In Section~5 we discuss the cosmological constraints on
${\tilde \nu}$ NLSP scenarios, and show that they are not severe. In Section~6
we discuss some signatures of metastable sneutrinos in different NUHM scenarios,
in particular those with different lepton flavours accompanying the ${\tilde \nu}$ NLSP.
Our conclusions are summarized in Section 7. Calculations of the three-body 
decay ${\tilde \nu} \to {\tilde G} \nu \gamma$ are described in an Appendix.

\section{Sneutrino NLSP Properties}

We assume that the GUT scale is the effective input scale at which the soft masses are specified, presumably via some gravity-mediated mechanism, and make the NUHM
assumptions that the gaugino masses are universal, as are the squark and slepton masses,
whereas the soft supersymmetry-breaking contributions to the Higgs scalar masses
are non-universal. We then calculate the
physical supersymmetric mass parameters at a low energy scale from the running
given by the renormalization-group equations (RGEs). We assume
that the right-handed neutrino supermultiplets, being singlets, get very large
Majorana masses, and therefore decouple from the low-energy effective theory \footnote{However,
if the decoupling energy scale is significantly below the GUT scale, the low-energy spectrum
may be affected \cite{Barger} through the RGEs. This effect is neglected here.}.
The sneutrino NLSP discussed in this paper is essentially the scalar partner  of some 
left-handed neutrino. The flavour of the ${\tilde \nu}$ NLSP is, however,  model-dependent, as we
discuss below.

\subsection{Sneutrino Mass}

In order to calculate the sneutrino mass, we first look at the RGEs of the
slepton sector~\footnote{Although we write the one-loop RGEs for simplicity, our calculations 
include two-loop contributions.}
\cite{Barger:1993gh,deBoer:1994he,Martin:1993zk}:
\begin{eqnarray}
\frac{d m_{\widetilde{L}_L}^2}{dt} &=& \frac{1}{8 \pi^2} (-3 g_2^2 M_2^2 -
          g_1^2  M_1^2 - 2 S) ,  \nnl
\frac{d m_{\widetilde{e}_R}^2}{dt} &=& \frac{1}{8 \pi^2} (-4 g_1^2 M_1^2 + 4 S) ,
         \nnl 
\frac{d m_{\widetilde{L}_{3L}}^2}{dt} &=& \frac{1}{8 \pi^2} (-3 g_2^2 M_2^2 -
          g_1^2 M_1^2 +
          h_\tau^2 ( m_{\widetilde{L}_{3L}}^2 + m_{\widetilde{\tau}_R}^2 + 
	  m_1^2 + A_\tau^2 ) - 2 S) , \nnl
\frac{d m_{\widetilde{\tau}_R}^2}{dt} &=& \frac{1}{8 \pi^2} (-4 g_1^2 M_1^2 +
          2 h_\tau^2 ( m_{\widetilde{L}_{3L}}^2 + m_{\widetilde{\tau}_R}^2 +
	  m_1^2 +  A_\tau^2 ) + 4 S) ,
\label{rges}
\end{eqnarray}
where
\begin{eqnarray}
S &\equiv& \frac{g_1^2}{4} ( m_2^2 - m_1^2 +
        2 ( m_{\widetilde{Q}_L}^2  - m_{\widetilde{L}_L}^2 - 2
	m_{\widetilde{u}_R}^2 +
	m_{\widetilde{d}_R}^2 + m_{\widetilde{e}_R}^2 ) \nonumber \\ && \, + \, 
          ( m_{\widetilde{Q}_{3L}}^2 - m_{\widetilde{L}_{3L}}^2 - 2
	  m_{\widetilde{t}_R}^2 
	  + m_{\widetilde{b}_R}^2 + m_{\widetilde{\tau}_R}^2 )) .
\label{defS}
\end{eqnarray}
The $S$ term vanishes and does not contribute in models
with universal soft masses such as the CMSSM. 
However, for non-universal models, $S$ can be far from zero and its contribution
to the RGEs can be significant. When $S = 0$, and assuming universal soft
breaking masses  for the $L$ and $R$ sleptons at the GUT scale, the $R$
slepton is lighter than the $L$ slepton at the weak scale.
However, since $S$ contributes in opposite ways for $L$ and $R$ sleptons,
if $S$ is large and negative, $\widetilde{L}_L$ could be lighter than
$\widetilde{L}_R$.  Furthermore,
there are additional $D$-terms,  
\bear
m^2_{\tilde{e}_L} &=& m^2_{\widetilde{L}_L} - \cos (2 \beta) m_Z^2 \left( \half -
\sin^2 \theta_W \right) , \nnl
m^2_{\tilde{\nu}_L} &=& m^2_{\widetilde{L}_L} + \cos (2 \beta) \half m_Z^2 ,
\eear
that split the sneutrino and charged-slepton masses. Since $\cos (2 \beta) < 0$ for 
$\tan \beta > 1$, the sneutrino is lighter than its charged-slepton partner.

Comparing the first two generations and the third generation, we see
that the tau sneutrino could be lighter than the electron and muon sneutrinos because of
the Yukawa terms in the RGEs. However, this is not always the case, due to the
fact that $m_1^2$ could be negative and large, and we display examples later
where the ${\tilde \nu}$ NLSP has electron or muon flavour.
The lighter stau mass is also suppressed by
off-diagonal terms in the mass matrix. Thus, depending on the model
parameters, either the tau sneutrino or the lighter stau might be lighter.

Therefore, a sneutrino could be the NLSP if $S$ is large and negative~\footnote{One could
also obtain a
light sneutrino within a supersymmetric SU(5) GUT with different soft masses for the
${\bf 10}$ and $\bar{\bf 5}$ multiplets~\cite{KKKM-2}. Another alternative is
within a gaugino-mediated supersymmetry breaking model, in which the Higgs masses
are again different from the other sfermion masses~\cite{BCKS}.}. 
We see from (\ref{defS})  that $S$ is negative when $m_2^2 - m_1^2 <
0$~\footnote{Assuming that these are the dominant terms in $S$, which is the
case for the NUHM that we consider here.},
which is not possible
in the CMSSM, for which $S = 0$ by assumption. We now study how this may occur
in the NUHM model, using the freedom that the Higgs soft
supersymmetry-breaking masses at the GUT scale are not necessarily equal to $m_0$, the
universal scalar mass for sleptons and squarks. 

The electroweak symmetry breaking conditions may be written in the form:
\begin{equation}
m_A^2 (Q) = m_1^2(Q) + m_2^2(Q) + 2 \mu^2(Q) + \Delta_A(Q)
\end{equation}
and
\begin{equation}
\mu^2 = \frac{m_1^2 - m_2^2 \tan^2 \beta + \frac{1}{2} \mz^2 (1 - \tan^2 \beta)
+ \Delta_\mu^{(1)}}{\tan^2 \beta - 1 + \Delta_\mu^{(2)}} \ ,
\end{equation}
where $\Delta_A$ and $\Delta_\mu^{(1,2)}$ are loop
corrections~\cite{Barger:1993gh,deBoer:1994he,Carena:2001fw} and $m_{1,2} \equiv
m_{1,2}(\mz)$~\footnote{Our convention is such that $H_{1,2} \equiv H_{d,u}$,
and $\tan \beta \equiv v_2/v_1$.}.  
The values of the
NUHM parameters at $Q$ are related to their values at $\mz$ through 
the known radiative corrections~\cite{Barger:1993gh,IL,Martin:1993zk} 
$c_1, c_2$ and $c_\mu$: 
\begin{eqnarray}
m_1^2(Q) &=& m_1^2 + c_1 \, , \nnl
m_2^2(Q) &=& m_2^2 + c_2 \, , \nnl
\mu^2(Q) &=& \mu^2 + c_\mu \, .
\end{eqnarray}
Solving for $m^2_1$ and $m^2_2$, one has
\begin{eqnarray}
m_1^2(1+ \tan^2 \beta) &=& m_A^2(Q) \tan^2 \beta - \mu^2 (\tan^2 \beta + 1 -
\Delta_\mu^{(2)} ) 
- (c_1 + c_2 + 2 c_\mu) \ttbt \nl - \Delta_A(Q) \ttbt 
- \frac{1}{2} \mz^2 (1 - \ttbt) - \Delta_\mu^{(1)} 
\label{m1}
\end{eqnarray}
and 
\begin{eqnarray}
m_2^2(1+ \tan^2 \beta) &=& m_A^2(Q) - \mu^2 (\tan^2 \beta + 1 +
\Delta_\mu^{(2)} )
- (c_1 + c_2 + 2 c_\mu) \nl
- \Delta_A(Q) + \frac{1}{2} \mz^2 (1 - \ttbt) + \Delta_\mu^{(1)} \ .
\label{m2}
\end{eqnarray}
The correction $\Delta_\mu^{(2)}$ is positive and generally of $O(0.1)$. 
From here we can see that there are two possible ways to get negative $S$ via
negative $m_2^2 - m_1^2$: the first is by using very large $\mu^2$, and the
second is by using very large $m_A^2$. 
If $m_A^2$ is relatively small while $\mu^2$ is very large,  
the Higgs masses-squared $m_{H_{1,2}}^2 \equiv m_{1,2}^2 + \mu^2$ may be
negative at the GUT scale, which could lead to a vacuum stability problem \cite{eoss3}. 

A weak-scale scalar mass-squared with $m^2(M_{GUT}) < 0$ generally produces
a vev of order the weak scale that disappears as the RGEs are run down to
the weak scale. Such negative masses-squared are not dangerous. It may happen, however,
that an instability occurs along some $F-$ and $D-$ flat direction.  In this case, 
a negative mass-squared may be large and still present at a renormalization scale, $Q \sim v$ \cite{fors}. 
Models in which the Universe becomes trapped in such non-Standard Model vacua are
clearly excluded~\footnote{Whether this occurs or not depends on the 
specific cosmological history during inflation \cite{thermal,eglos}.}.
This possibility has been studied in the NUHM along the $H_1-H_2$ and $H_2 - L$ flat
directions of the MSSM~\cite{eglos}.
We delineate below regions in the parameter plane where these vacua may be problematic. 

\subsection{Sneutrino Lifetime}

In the GDM scenario used here, the sneutrino NLSP would eventually decay into
the gravitino, and the dominant decay channel is the two-body decay
\beq
\snu \to \widetilde{G} + \nu \ ,
\eeq
with the decay rate
\beq
\Gamma_{\rm 2b} = \frac{1}{48 \pi} 
\frac{\msnu^5}{M_{\rm Pl}^2 \mgrav^2} 
\left( 1 - \frac{\mgrav^2}{\msnu^2} \right)^4    ,
\label{2bodylife}
\eeq
where $\mgrav$ is the gravitino mass and $M_{\rm Pl}$ is the Planck mass: $M_{\rm
Pl} = 1/\sqrt{8 \pi G_N} \simeq 2.4 \times 10^{18}$~GeV. 

\begin{figure}[ht]
\begin{center}
\mbox{\epsfig{file=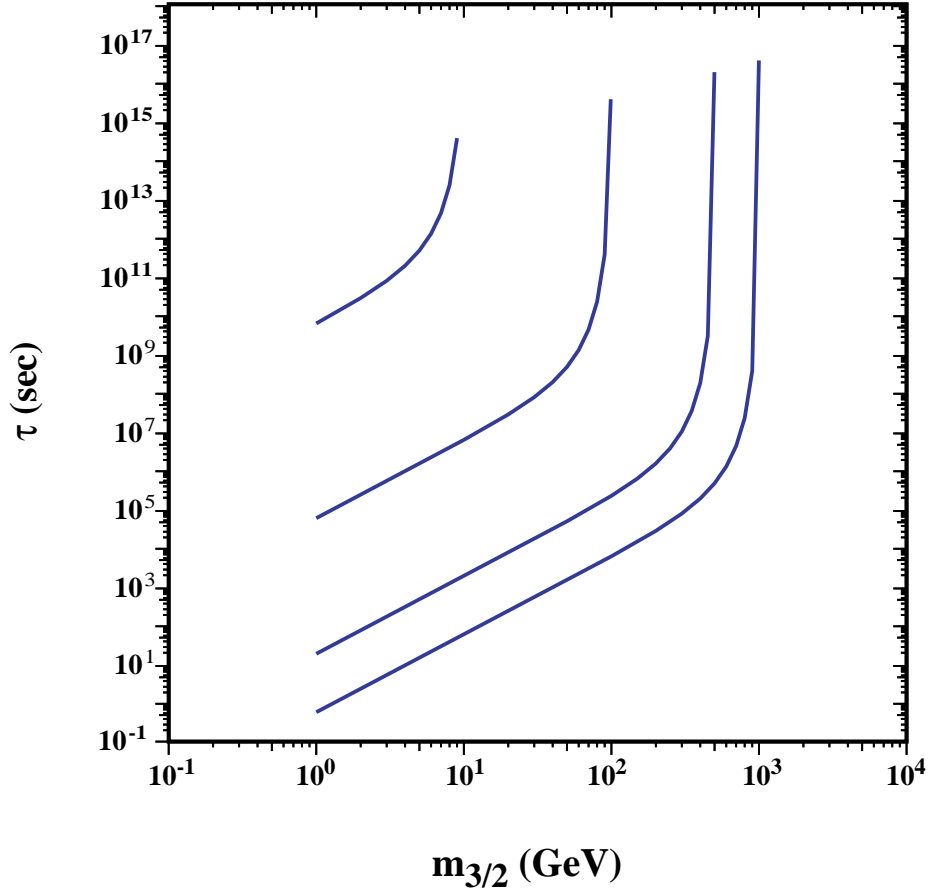,height=12cm}}
\end{center}
\caption{\label{fig:2body}\it
The sneutrino NLSP lifetime as a function of $\mgrav$ for $\msnu = 10, 100,
500$ and $1000$~GeV (top to bottom).  
}
\end{figure}

We plot in Fig.~\ref{fig:2body} the sneutrino lifetime, $\tau_{\tilde{\nu}}
\simeq 1/\Gamma_{\rm 2b}$, as a function of the
gravitino mass for $\msnu = 10, 100, 500$, and 1000~GeV respectively. 
Note that we plot the lifetime only for 
$\Delta m \equiv \msnu - \mgrav \ge 1$~GeV. Clearly, a smaller mass gap would yield an even
longer lifetime. We see that
the sneutrino lifetime could be less than 1~second only when $\msnu$ is large,
or the gravitino mass is (much) less than 1~GeV. On the other hand, if the
mass gap is small, the sneutrino lifetime can be very long, potentially even longer than
the age of the Universe, which is ${\cal O}(10^{17})$~s.
However, there are cosmological and astrophysical constraints on the possibility
of a sneutrino with lifetime longer than the age of the Universe at recombination that
we discuss in more detail later. 
 
\section{Cosmological Sneutrino Density}

Assuming thermal equilibrium in the early Universe, one can calculate the
sneutrino relic density after decoupling but before its decay into the gravitino. 
This is done by the usual use of the Boltzmann equation and calculation of the
sneutrino annihilation and coannihilation cross sections. The calculations
are identical to those required to calculate the relic sneutrino density if
it is the LSP. 

The possible sneutrino-pair annihilation two-body final states are the
following~\cite{ourNUHM2}:

\begin{center}
\begin{tabular}{c|l}
Initial State & Final States\\ 
\hline
~ & ~ \\
${\tilde \nu}_i {\tilde \nu}_i^\ast$ & $f \bar{f}, W^+ W^-, Z Z, hZ, hA, HZ, HA,
hh, hH, HH, AA, AZ, H^+ H^-, W^+ H^-, H^+ W^-$ \\
${\tilde \nu}_i {\tilde \nu}_i$  & $\nu_i \nu_i$ 
\\ 
\end{tabular}
\label{table:anni}
\end{center}

If the soft masses for the sfermions are universal as assumed here, the electron sneutrino is
always degenerate with the muon sneutrino, and the tau-sneutrino mass might be nearby.
In the NUHM case that we consider here,
there could also be other sparticles that are almost degenerate with the 
sneutrinos, such as the lightest neutralino and chargino, and charged
sleptons. We list below the corresponding coannihilation processes and their
possible final states:
\begin{itemize}
\item Coannihilation with other sneutrino flavours:
\begin{center}
\begin{tabular}{c|l}
Initial State & Final States\\ 
\hline
~ & ~ \\
${\tilde \nu}_i {\tilde \nu}^\ast_j$ & $\nu_i \bar{\nu}_j, \ell_i \bar{\ell}_j$ 
\\
${\tilde \nu}_i {\tilde \nu}_j$  & $\nu_i \nu_j$ \\
\end{tabular}
\label{table:coanni1}
\end{center}

\item Coannihilation with charged sleptons:
\begin{center}
\begin{tabular}{c|l}
Initial State & Final States\\ 
\hline
~ & ~ \\
${\tilde \ell}_i {\tilde \nu}^\ast_i$ & $f \bar{f}^\prime, h H^-, H H^-, A H^-,
h W^-, H W^-, Z W^-, \gamma W^-, W^- A$ 
\\
${\tilde \ell}_i {\tilde \nu}^\ast_j$  & $\ell_i \bar{\nu}_j$ \\
${\tilde \ell}_i {\tilde \nu}_j$  & $\ell_i \nu_j, \nu_i \ell_j$ \\
\end{tabular}
\label{table:coanni2}
\end{center}
In the first line above, the coannihilation is between a sneutrino and a charged
slepton of the same generation, whereas in the second and third lines they are not
necessarily from the same generation.

\item Coannihilation with the lightest neutralino:
\begin{center}
\begin{tabular}{c|l}
Initial State & Final States\\ 
\hline
~ & ~ \\
$\chi {\tilde \nu}$ & $\nu Z, \nu h , \nu H , \nu A, \ell^- W^+, \ell^- H^+$ 
\\
\end{tabular}
\label{table:coanni3}
\end{center}
and similarly for $\snu^\ast$. 

\item Coannihilation with the lighter chargino:
\begin{center}
\begin{tabular}{c|l}
Initial State & Final States\\ 
\hline
~ & ~ \\
$\chi^- {\tilde \nu}$ & $\ell \gamma, \ell Z, \ell h, \ell H, \ell A, \nu W^-
, \nu H^-$ 
\\
$\chi^+ {\tilde \nu}$ & $\nu W^+, \nu H^+$ 
\\
\end{tabular}
\label{table:coanni4}
\end{center}
and similarly for $\snu^\ast$. 

\end{itemize}

In our scenario, the sneutrino NLSP eventually decays into the gravitino before the
current epoch. Consequently, the gravitino relic density is related to the sneutrino density
before its decay by
\beq
\Omega_{\tilde{G}} h^2 = \frac{\mgrav}{\msnu} \Omega_{\tilde{\nu}} h^2 +
\Omega_{\tilde{G}}^{\rm T} h^2  ,
\label{eq:relic}
\eeq
where $\Omega_{\tilde{G}}^{\rm T}$ is the contribution to the gravitino density from thermal
production after reheating, which is sensitive to the unknown reheating temperature
$T_R$. We do not discuss this contribution here. The only constraint we
impose is that the contribution to the gravitino relic density arising from
sneutrino decay does not exceed the value
suggested by WMAP~\cite{wmap} and other observations: 
\beq
\Omega_{\rm DM} h^2 = 0.1099 \pm 0.0062  \, .
\eeq
Hence, we require that the first term on the right-hand side of (\ref{eq:relic})
should not exceed $\sim 0.1223$ (the 2-$\sigma$ upper limit).
Because of the scaling by the mass ratio $\mgrav/\msnu$, even a large
sneutrino density after decoupling could still be compatible with the dark
matter constraint if $\mgrav \ll \msnu$.  

In this case, we must check whether gravitinos are non-relativistic at the time
structure formation begins, roughly at $t_s \simeq 5 \times 10^{11}$~s, in which
case they act in the same way as conventional cold dark matter.
If $m_{3/2} \ll \msnu$, $E_{3/2}/m_{3/2} \simeq \msnu/2m_{3/2}$ 
and $E_{3/2}$ scales subsequently as $(\tau_{\tilde{\nu}}/t)^{1/2}$,
where $\tau_{\tilde{\nu}}$ is the sneutrino lifetime.  We can use Fig.~\ref{fig:2body} to estimate
whether or not gravitinos will behave as cold dark matter. For example,
at the left-most point of the lowest curve, gravitinos are produced relativistically
with $E_{3/2}/m_{3/2} \sim 1000$, but the decay occurs so early that they become
non-relativistic well before structure formation begins. The same is also true for the
two middle curves in Fig.~\ref{fig:2body}. Only for the left
part of the topmost curve (when $\msnu = 10$ GeV) is there a potential problem.
However, in this case the sneutrino relic density is generally already small
in the model considered here.

\section{NUHM Parameter Space}

In the CMSSM, the values of $m_A$ and $\mu$ are determined by the electroweak
vacuum conditions for any given input values of $m_{1/2}, m_0, A_0$ and $\tan \beta$.
However, the constraints on $m_A$ and $\mu$ are relaxed in the NUHM, with
the values of these parameters being related to the degrees of non-universality
assumed for the Higgs soft masses $m_1^2$ and $m_2^2$. A general discussion of
the 
parameter space of the NUHM was given in~\cite{ourNUHM1,ourNUHM2,baerNUHM,eosk4}, 
which we use here as a
starting point for our discussion. Regions of the NUHM parameter space in which a
sneutrino is the lightest spartner of any Standard Model particle were
identified in~\cite{ourNUHM2}, see for example the dark blue shaded regions in
Figs.~2, 3, 4, 6, 8 and 9 of that paper. There, it was assumed that
the LSP is the lightest
neutralino $\chi$, with the gravitino assumed to be heavy, so that these
light-sneutrino regions were disallowed. However, in this paper we assume that
the gravitino is the
LSP, so the viability of these light-sneutrino regions must be re-evaluated.
We focus our discussion here on $(\mu, m_A)$ planes of
the types shown in Figs.~4 and 6 of~\cite{ourNUHM2}, where the light-sneutrino
region moves to lower $|\mu|$ as $m_A$ increases.

\begin{figure}
\begin{center}
\mbox{\epsfig{
file=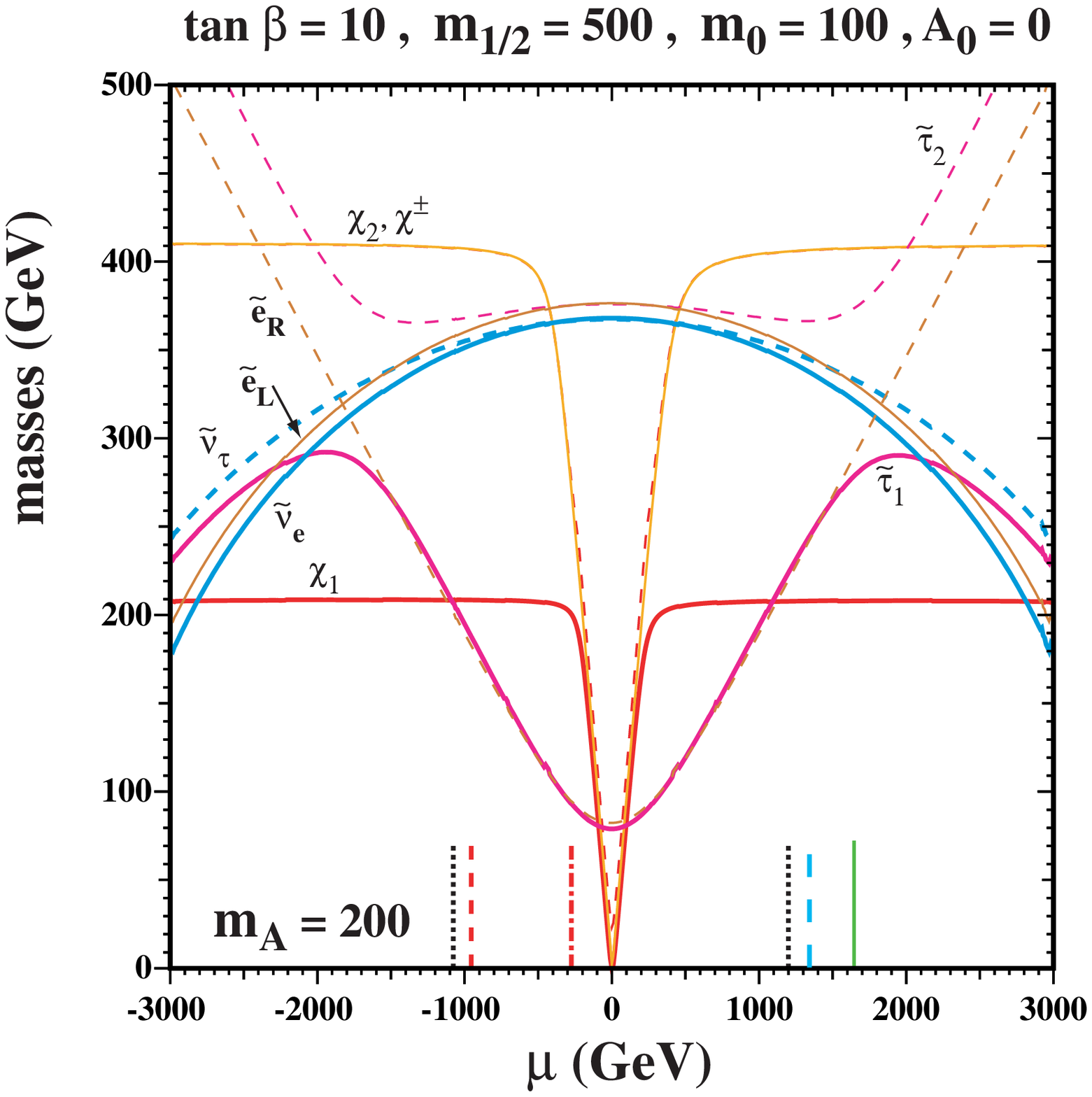,height=8cm}}
\mbox{\epsfig{
file=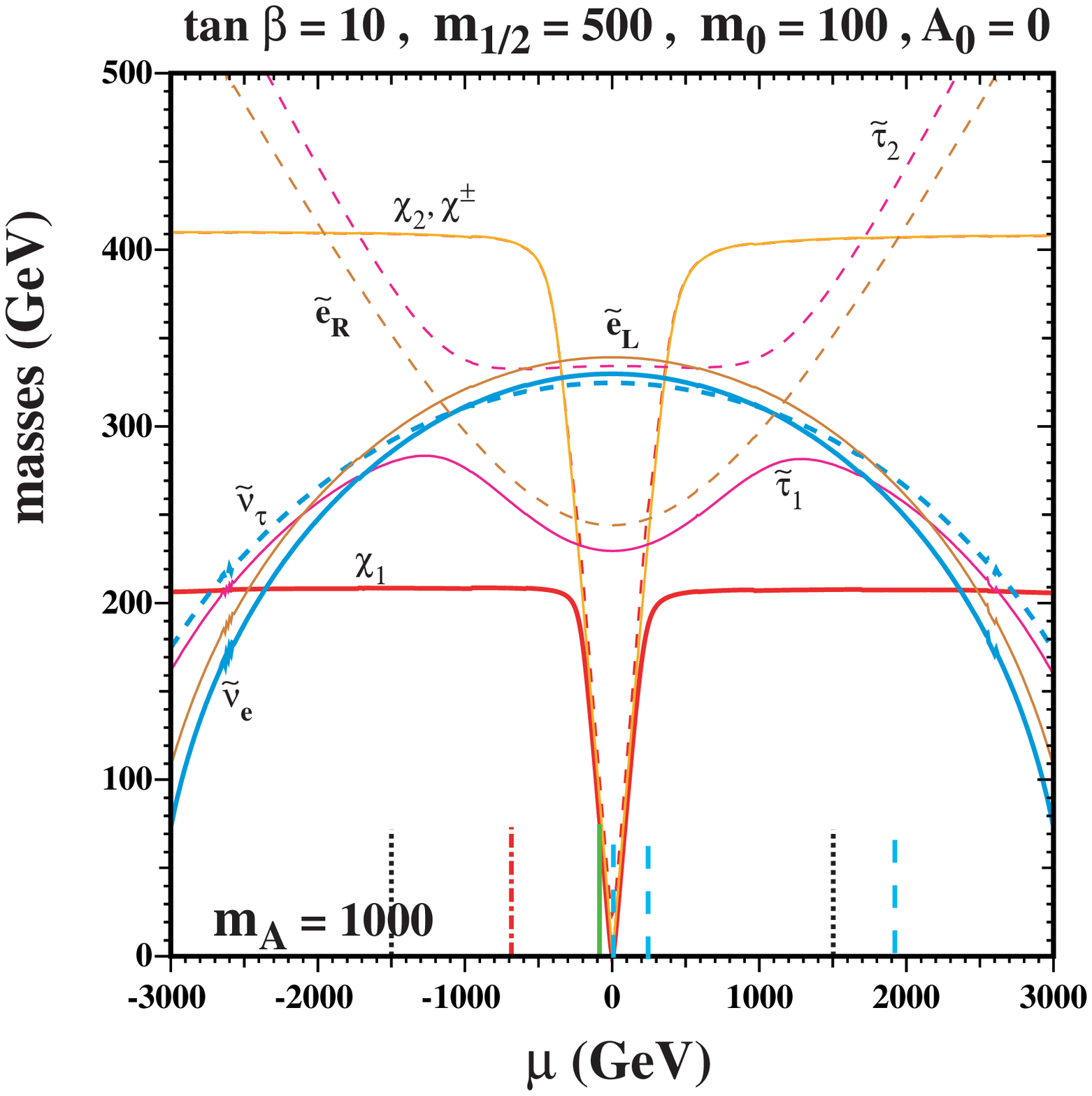,height=8cm}}
\end{center}
\begin{center}
\mbox{\epsfig{
file=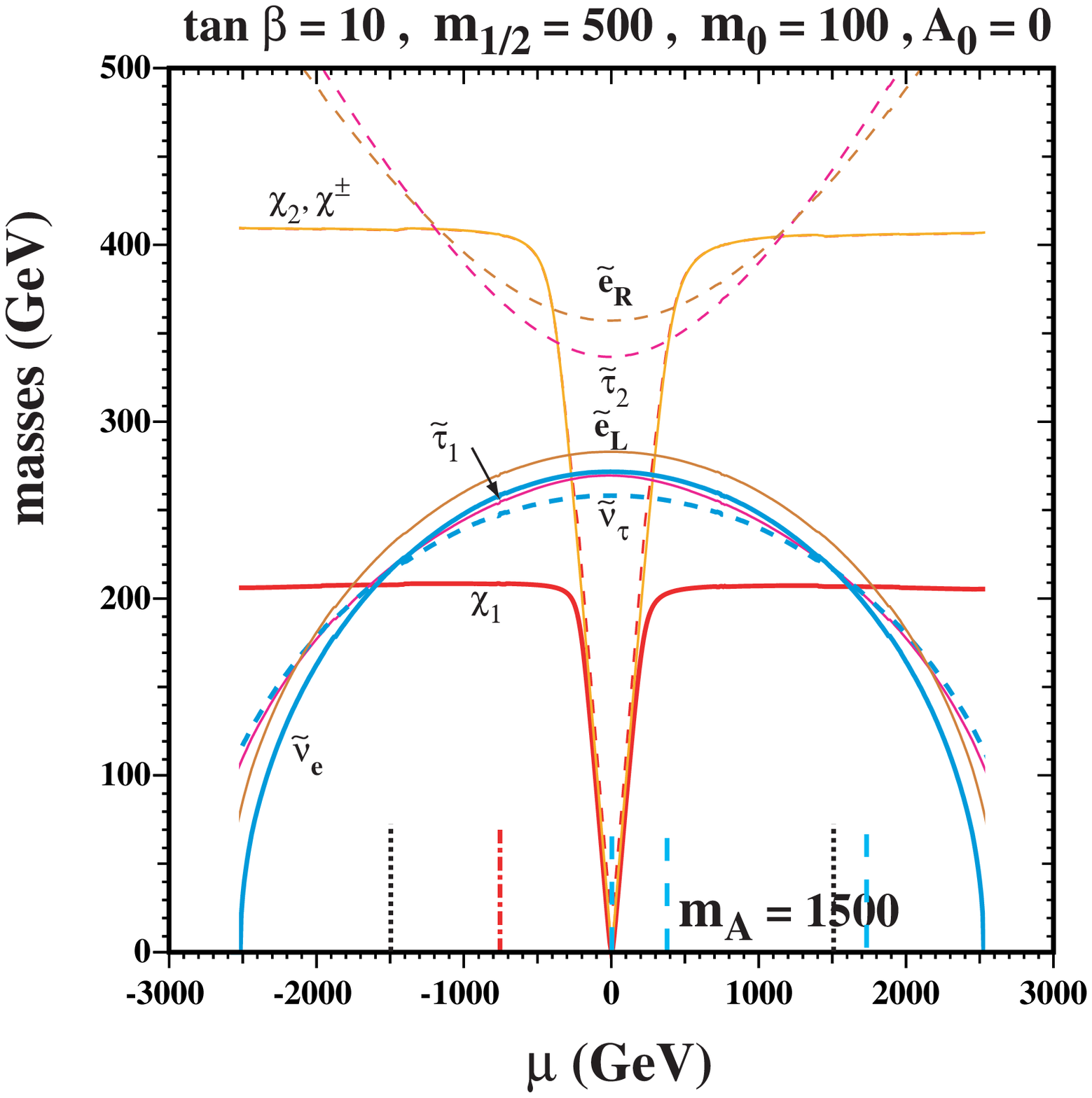,height=8cm}}
\mbox{\epsfig{
file=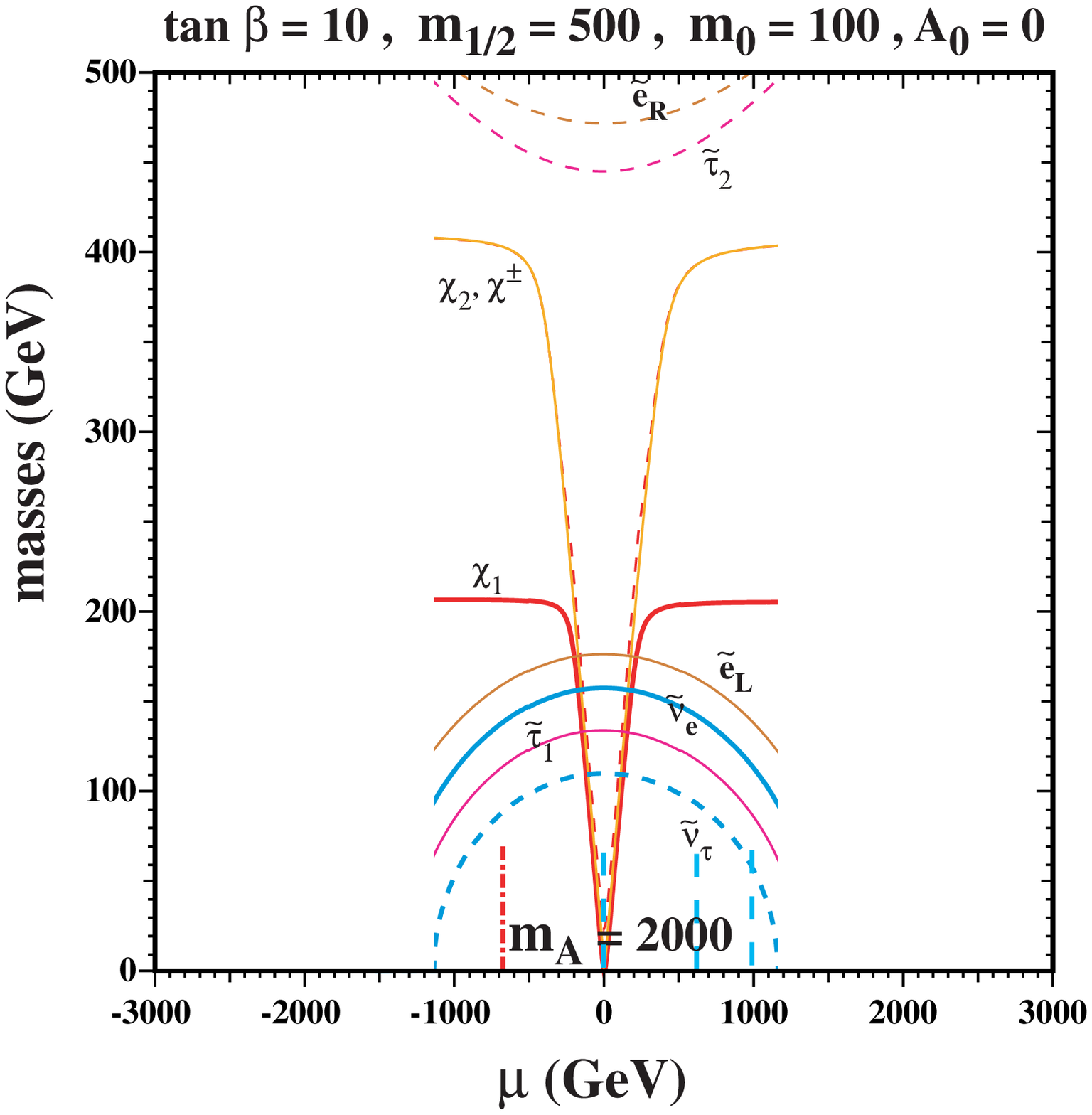,height=8cm}}
\end{center}
\caption{\label{fig:massvsmu} \it Sparticle masses as functions of $\mu$
for $\tan \beta = 10$, $m_{1/2} = 500$~GeV, $m_0 = 100$~GeV, $A_0 = 0$, $m_t =
172.6$~GeV, $m_b(m_b)^{\overline{\rm MS}} =4.25$~GeV, and $m_A =$ (a) 200~GeV,
(b) 1000~GeV, (c) 1500~GeV and (d) 2000~GeV, respectively. In panels (c) and (d),
the sparticle lines are truncated at larger $|\mu|$ where some sneutrino becomes tachyonic.
Constraints are
represented by vertical lines: black dotted for the GUT constraint (larger
$|\mu|$ is excluded); red dot-dashed shows the Higgs mass 
contour at $m_h = 114.4$ GeV, while the constraint using the LEP
likelihood function convolved with theoretical uncertainties in the Higgs mass
(computed here using {\tt FeynHiggs} \cite{fh}) is shown by the red dashed line; 
the $(g-2)_\mu$ constraint (described in the text)
 is shown by the light blue long dashed lines; and solid green for the $b \to s \gamma$ constraint (smaller $\mu$
is excluded). } 
\end{figure}

In general, keeping some sneutrino species light favours small values of
$m_{1/2}$ and $m_0$. However, there is an important lower limit on $m_{1/2}$, 
in particular, due to the LEP lower bound on the mass of the lightest MSSM Higgs boson. 
A sneutrino NLSP region may be found by choosing a moderate value of 
$m_{1/2} = 500$~GeV while keeping $m_0$ relatively small, e.g., $m_0 = 100$~GeV.  
The resulting masses of the sparticles are shown in 
Fig.~\ref{fig:massvsmu} assuming $\tan \beta = 10$ and $A_0 = 0$, 
for various values of $m_A = 200, 1000, 1500$ and 2000~GeV, 
in panels (a), (b), (c), and (d), respectively~\footnote{We use $m_t =172.6$~GeV for
our analysis~\cite{mt-2008}.}.

As noted earlier, in regions of the NUHM parameter plane (particularly
when $\mu$ and $m_A$ are large), the masses-squared of the Higgs and 
left-sleptons have a tendency to run down to negative values at the GUT scale.
This allows for the possibility that large scale vevs be excited along
the $H_1 - H_2$ or $H_2 - L$ flat directions. One  expects that these
flat directions are lifted by some effective operator at or above the GUT scale. 
The vev along the flat direction is sensitive to the fundamental scale
associated with this operator and clearly grows as that scale is increased above
the GUT scale. The reliability of this high-scale vacuum depends also on
the one-loop corrections to the scalar potential.  This sensitivity can
be characterized by the ratio of the tree-level vev to the renormalization
scale, $Q_0$, at which the vev disappears (i.e., the masses-squared go positive).
Here we adopt the most conservative set of assumptions, namely that
the flat directions are lifted at the GUT scale and that the vev must be of order 
$Q_0$ ($\epsilon = 1$ in the notation of \cite{eglos}). This preserves the largest
volume of the NUHM parameter space.  Of course the GUT constraint itself
is cosmology-dependent, and may not be important if the Universe starts out
in the weak scale vacuum after inflation.  For more details on this constraint see~\cite{eglos}.

In the case of small $m_A = 200$~GeV, shown in panel (a) of
Fig.~\ref{fig:massvsmu}, we see that the GUT stability constraint (represented
by a couple of vertical black dotted lines) allows only small $|\mu| \lappeq
1.1$~TeV, far from the
sneutrino NLSP region which appears when $|\mu| \gappeq 2800$~GeV.  We also see
that the $b \to s \gamma$ constraint (solid green line) allows only  $\mu
\gappeq 1600$~GeV and the Higgs mass constraint (dashed red) allows only $\mu
\gappeq -1000$~GeV for this small value of $m_A$. Note that a Higgs mass of
114.4 GeV
is found around $\mu \simeq -300$ GeV (dot-dashed vertical line) for this value of $m_A$,
 but current theoretical and experimental
uncertainties only provide for the weaker bound shown by the dashed line.
The anomalous magnetic moment of the muon, $(g-2)_\mu$,
is reconciled with experiment at the 95\% CL for $\mu \gappeq 1300$ GeV,
as shown by the light blue long-dashed line~\footnote{We assume that the
deviation of $(g-2)_\mu/2$
from the standard model is between 10.7 to $44.3 \times
10^{-10}$, the 2 $\sigma$ range according to~\cite{Davier:2007ua}.}.
Therefore, we do not have an
allowed region in panel (a), assuming the GUT constraint holds. For larger $m_A$,
these
constraints become more relaxed, and an allowed sneutrino NLSP region emerges.

For $m_A = 1000$~GeV, shown in panel (b) of Fig.~\ref{fig:massvsmu}, the GUT
stability constraint allows $|\mu| \lappeq 1.5$~TeV, the $b \to s \gamma$
constraint allows $\mu \gappeq 0$, and the Higgs constraint is essentially unimportant,
as a Higgs mass greater than 114.4 GeV occurs at $\mu \gappeq
- 700$~GeV.  The $(g-2)_\mu$ constraint is satisfied in two regions: $0 < \mu < 250$ GeV and 
$\mu \gappeq 1900$ GeV.  While the GUT constraint is relaxed, the sneutrino LSP region
still requires $2.4$~TeV $\lappeq |\mu|$,
where we have degenerate ${\tilde \nu_{e,\mu}}$ NLSPs.
We note that around $\mu = 2.5$~TeV several other sparticles are only slightly heavier than
the ${\tilde \nu_{e,\mu}}$, including the lightest neutralino $\chi$, the
${\tilde e_L}$ and  ${\tilde \mu_L}$, the lighter ${\tilde \tau}$ and the
${\tilde \nu_{\tau}}$. Thus, all these sparticles must be included in
coannihilation calculations of the ${\tilde \nu_{e,\mu}}$ abundance. For larger
$\mu$, only ${\tilde e_L}$ and  ${\tilde \mu_L}$ masses stay close to the NLSP
mass, while the others get larger mass gaps.

When $m_A$ is increased to 1500~GeV, as shown in panel (c) of
Fig.~\ref{fig:massvsmu},  the GUT constraint remains at 
$|\mu| \lappeq 1.5$~TeV, very close to the region when the 
sneutrino  is the NLSP, which extends from $\mu \simeq
1.6$~TeV to $\simeq 2.5$~TeV. The lightest sparticles are again the ${\tilde
\nu_{e,\mu}}$, with the $\chi, {\tilde e_L}, {\tilde \mu_L}, {\tilde \tau_1}$
and ${\tilde \nu_{\tau}}$ again slightly heavier. In this case, a theoretical
upper limit on $|\mu|$ arises when the ${\tilde \nu_{e,\mu}}$ become 
tachyonic~\footnote{Here and in panel (d), we truncate all the other sparticle
lines at this boundary of the tachyonic region.
Slightly more stringent upper limits on $|\mu|$ come from the lower limits on $m_{\tilde \nu}$
provided by LEP~\cite{efos,rpp}.}.
In this case, the Higgs constraint requires only that $\mu \gappeq -1300$ GeV,
and $(g-2)_\mu$ is satisfied when $0 < \mu \lappeq 400$ GeV or $\mu \gappeq 1700$
GeV.  The constraint from $b \to s \gamma$ is unimportant at this value of $m_A$.

Finally, in panel (d) of Fig.~\ref{fig:massvsmu} we display the sparticle masses
for $m_A = 2000$~GeV. In this case, the allowed sneutrino NLSP is the ${\tilde
\nu_{\tau}}$, for 200~GeV $\lappeq |\mu| \lappeq 1.1$~TeV. The $\chi$ becomes the
NLSP for smaller $|\mu|$ in the Higgsino region, and upper limits are provided by
the LEP lower limits discussed above. For this value of $m_A$, the differences in mass
between the ${\tilde \nu_{\tau}}$  and the heavier sparticles are relatively
large. Neither the Higgs mass nor $b \to s \gamma$ provide a constraint, while
$(g-2)_\mu$ requires $0 < \mu \lappeq 600$ GeV or $\mu \gappeq 1000$ GeV with
the later has $m_{\tilde \nu}$ less than the LEP limit.

In these plots, we find that sneutrino NLSP has relic density of order
$O(10^{-3})$ which is well below the WMAP limit. 
This means that most of the gravitino dark matter
must be produced by some other sources, e.g., by reheating.  
We take coannihilation effects into account for the relic density
calculation. However, in contrast with the neutralino LSP case, coannihilations 
do not always reduce the final relic density.
Sneutrino coannihilation with the lightest neutralino would indeed generally increase the
relic density, while that with charged sleptons might reduce it.   In the former (latter) case,
the effective sneutrino cross section is averaged with the weaker (stronger) annihilation
cross section of neutralinos (charged sleptons).  
The relatively small relic density of the sneutrino compared to that of the
neutralino can be attributed generically to the fact that the sneutrino is a scalar particle,
rather than a Majorana fermion. 
In Fig.~\ref{fig:massvsmu}(d), for example, since the mass gaps with other
sparticles are relatively large, the coannihilation effects are not
maximal, but the relic density is still
small.

\begin{figure}
\begin{center}
\mbox{\epsfig{
file=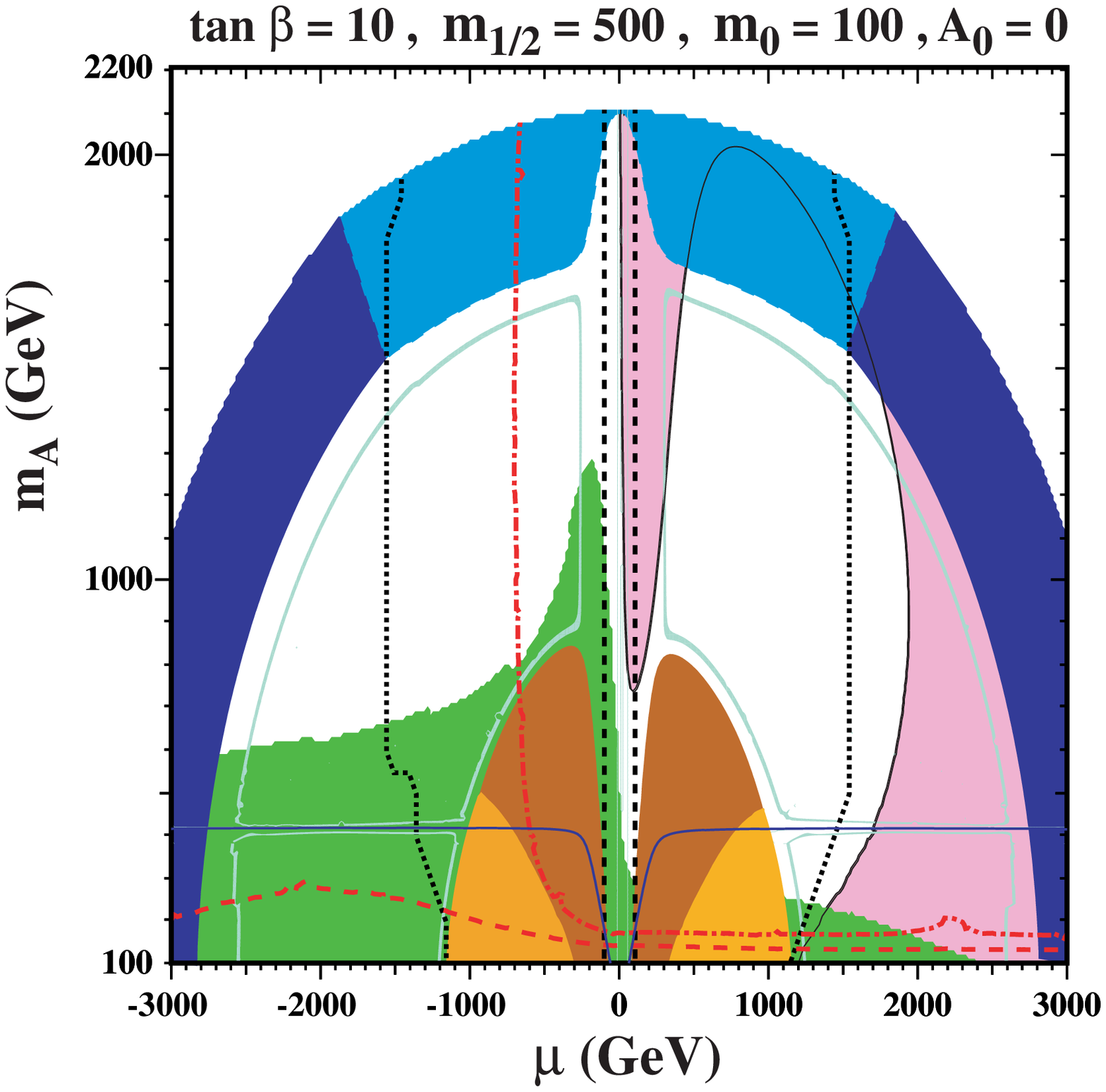,height=8cm}}
\mbox{\epsfig{
file=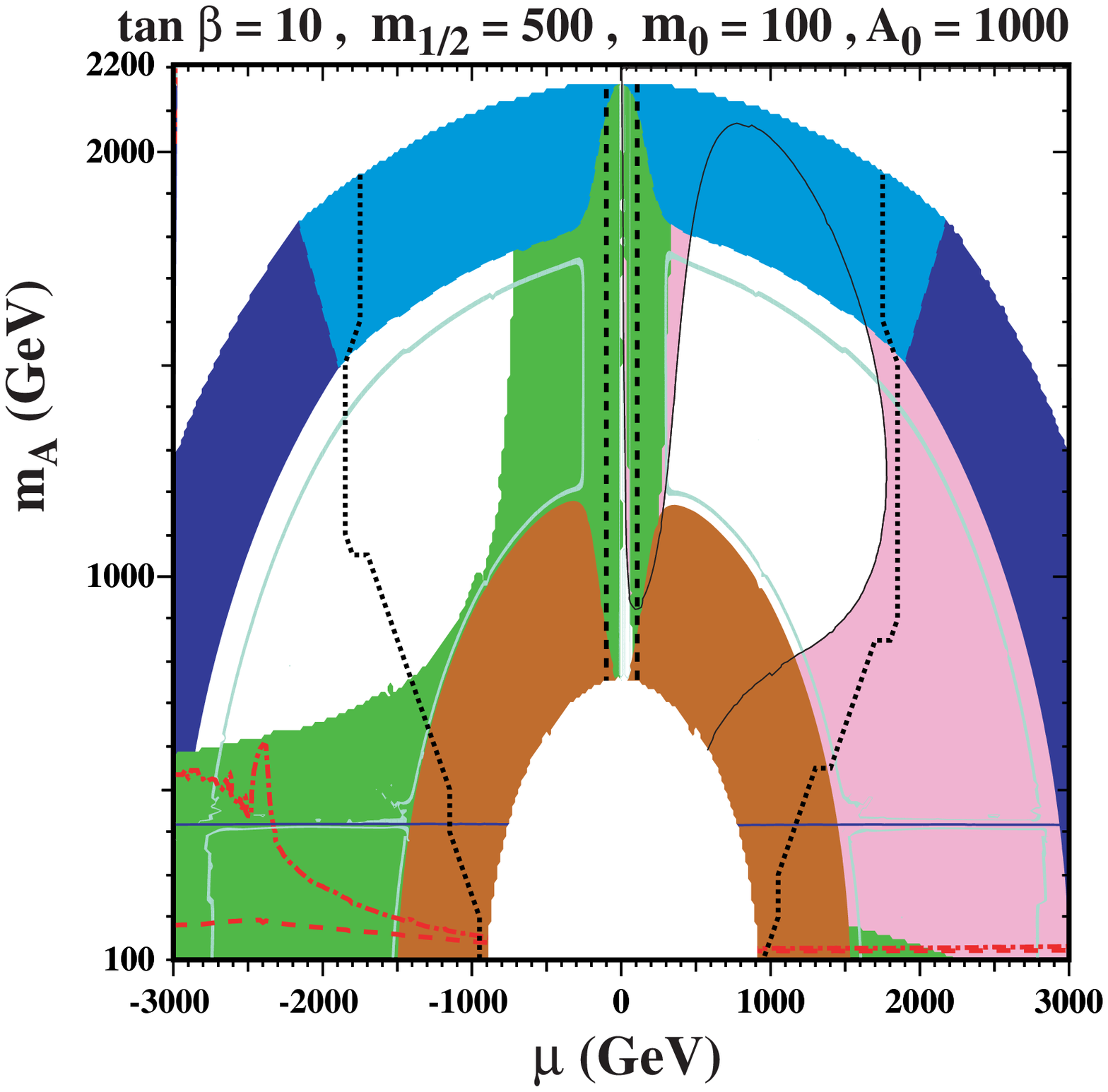,height=8cm}}
\end{center}
\begin{center}
\mbox{\epsfig{
file=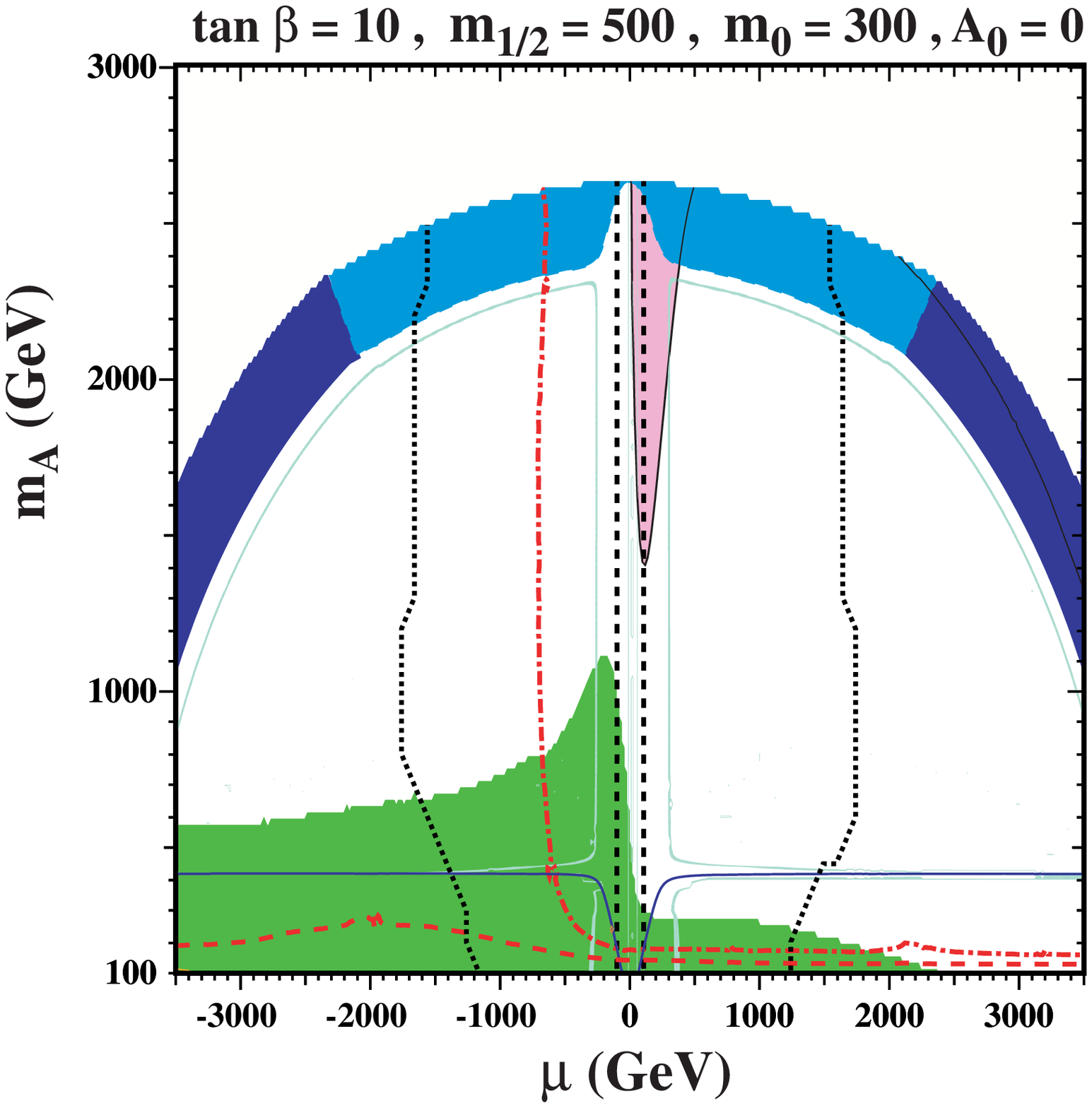,height=8cm}}
\mbox{\epsfig{
file=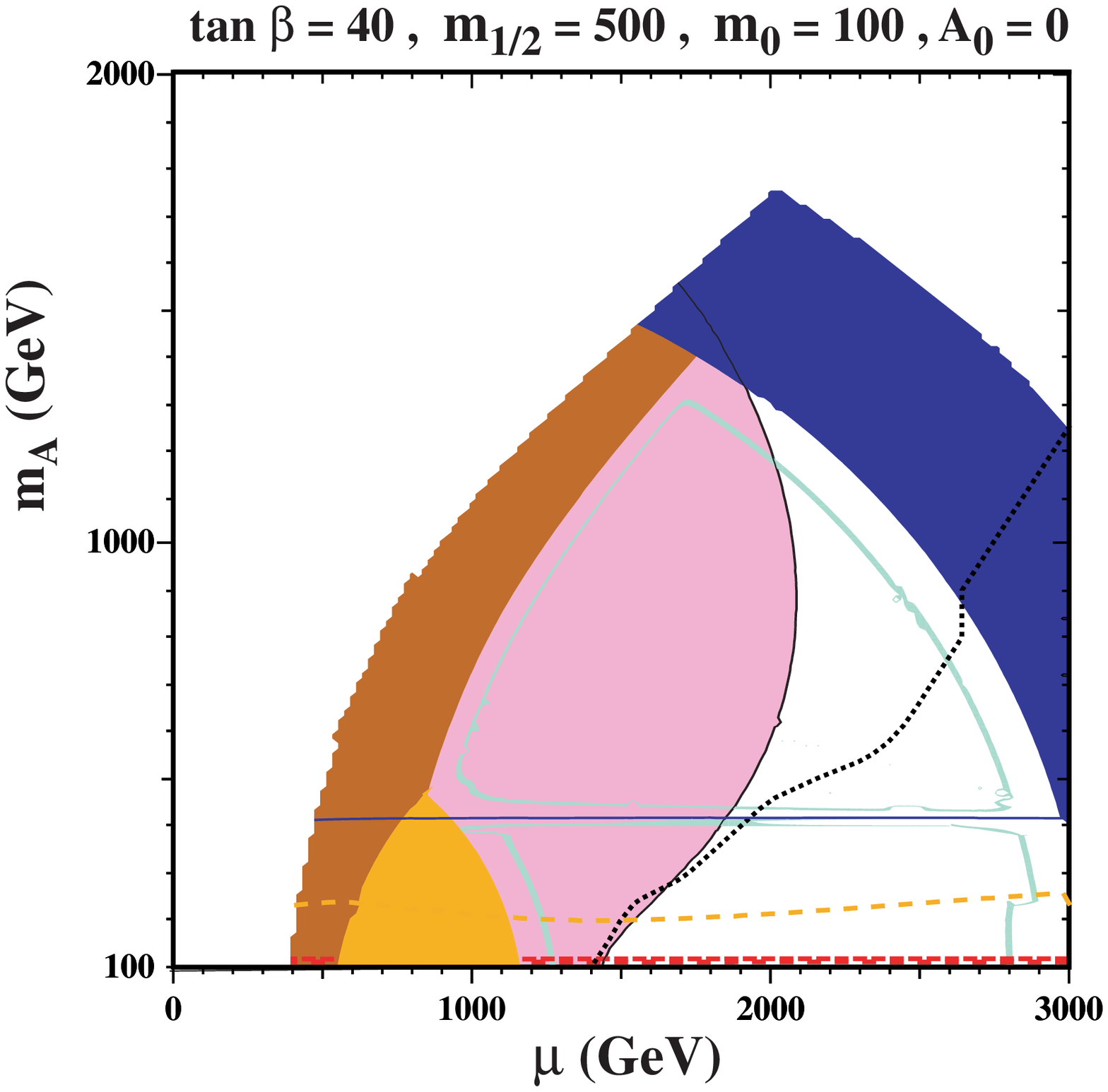,height=8cm}}
\end{center}
\caption{\label{fig:mumA} \it 
Some $(\mu, m_A)$ planes in the NUHM for (a) $\tan \beta = 10$, $m_{1/2} = 500$~GeV, 
$m_0 = 100$~GeV, and $A_0 = 0$; (b) $\tan \beta = 10$, $m_{1/2} = 500$~GeV, 
$m_0 = 100$~GeV, and $A_0 = 1000$; (c) $\tan \beta = 10$, $m_{1/2} = 500$~GeV, 
$m_0 = 300$~GeV, and $A_0 = 0$; (d) $\tan \beta = 40$, $m_{1/2} = 500$~GeV, 
$m_0 = 100$~GeV, and $A_0 = 0$. In each case, we used $m_t = 172.6$~GeV and 
$m_b(m_b)^{\overline{\rm MS}} =4.25$~GeV. Contours and shading are described in the text.} 
\end{figure}

To get a more comprehensive view of the NUHM parameter space, 
we display in Fig.~\ref{fig:mumA} some contour plots in selected
$(\mu, m_A)$ planes. In panel (a), we choose 
$\tan \beta = 10$, $m_{1/2} = 500$~GeV, 
$m_0 = 100$~GeV and $A_0 = 0$. This panel therefore includes and extends
the specific examples
shown in Fig.~\ref{fig:massvsmu}. We plot the regions where the lighter
stau, the right selectron, some sneutrino or the lightest neutralino is the NLSP. 
At large $|\mu|$, the electron-sneutrino is the NSLP and is shown by the regions
shaded dark blue. At large $m_A$ and smaller $|\mu|$, the NLSP becomes the
tau-sneutrino (shaded light blue).  Above these regions, the white area corresponds to an unphysical
region where one or more of the sparticles has a negative mass-squared at the weak scale.
Below these regions, in most of the area, it is the lightest neutralino which is the NLSP.
At lower $m_A$ and relatively small $|\mu|$, we see regions where the
lighter stau  (shaded brick red), or the right selectron (shaded orange) is the NLSP.
At even smaller $|\mu|$ the NLSP is a higgsino-like neutralino. 
The narrow turquoise
shaded region is that in which $\Omega_{\rm NLSP} h^2$ is within two $\sigma$ of the WMAP
value. (Recall that, with the gravitino as the LSP, this is not the dark matter relic
density.)
Regions surrounded by the strip (which have higher $\Omega_{\rm NLSP} h^2$)
might also be permitted if $\mgrav \ll m_{\rm NLSP}$, and regions not surrounded by the strip 
(which have lower $\Omega_{\rm NLSP} h^2$) would certainly be permitted by the
dark matter constraint. As one can see these regions track very closely the
degeneracy
lines between the neutralino and one of the four sparticles where the relic density is controlled 
by coannihilations or the funnel region  where 
$2 m_\chi \simeq m_A$ (The contour $2 m_\chi = m_A$ is shown by the thin blue
line.).

The Higgs mass contour of 114.4 GeV (red dot-dashed line) 
excludes $\mu \lappeq -700$~GeV. However, if one uses a likelihood analysis for the Higgs mass
and allows for theoretical uncertainties in the calculation of $m_h$, this
constraint is relaxed  and most of the displayed region is allowed, as shown by red dashed line. 
The $b \to s \gamma$ constraint (shaded green) excludes small $m_A$ and prefers
positive $\mu$. However, this constraint is unimportant in this panel at large
$m_A$. The GUT constraint is represented by a black dotted line: it excludes
large $|\mu|$ and essentially all of the electron-sneutrino NLSP region for this
set of parameters.
Finally, the light pink shaded region bordered by a black solid line represents the
region favored by $(g-2)_\mu$. The vertical dashed black lines correspond to the chargino mass
contours of 104 GeV and exclude very small values of $|\mu|$.

In panel (b), we display a case with $A_0 = 1000$~GeV, with the other parameters 
chosen to be the same as in (a). We see that the stau NLSP region becomes bigger and there
is no longer a right-selectron NLSP region.  In the white region
interior to the stau NLSP region, the stau has gone tachyonic at the weak scale, which
{\it is} problematic. We also see that
 the $b \to s \gamma$ constraint becomes stronger, especially for negative $\mu$. 
However, the qualitative features of the sneutrino NLSP band at large $|\mu|$ and/or
$m_A$ are similar. This feature is also retained in panel (c), where the larger 
value $m_0 = 300$~GeV is chosen, and also in panel (d), in which a larger
value $\tan \beta = 40$ is chosen. 
In panel (c) (which now extends to higher values of $m_A$), there is no longer
a charged NLSP.  In panel (d), we see again a region with
a right selectron NLSP.  The GUT constraint in this case constrains only the lower
right corner of the plane shown, and allows part of the electron-sneutrino
region. The $B_s \to \mu^+ \mu^-$ constraint (orange
dashed line) excludes a region with small $m_A$ \cite{bmm2}. 

We conclude that the possibility of a sneutrino NLSP
is quite generic in the NUHM, and certainly not much less plausible than the lighter stau.
This is in contrast to the CMSSM, where a stau NLSP is a generic feature at large $m_{1/2}$
and small $m_0$, but there is no possibility of a sneutrino NLSP.

\section{Cosmological Constraints on a Sneutrino NLSP}

The cosmological impact of a long-lived sneutrino depends on its
lifetime~\footnote{See~\cite{KKKM-1} and references therein.}. If the
sneutrino decays during or after Big-Bang Nucleosynthesis (BBN), it could alter predictions for the
light-element abundances. If the sneutrino decays around or after the time of
recombination, it could distort the blackbody spectrum of the CMB. If the
sneutrino decays at a very late time, its effect might be seen on the diffuse
neutrino and photon spectra. The production of relativistic neutrinos by
sneutrino decays could also change the equation of state and therefore the
evolution history of the Universe~\cite{GongChen}.

Although the sneutrino is neutral, and its dominant two-body decay channel produces only
a neutrino and the gravitino, which are also neutral, there could still be a significant
effect on BBN if the sneutrino decays during or round after the time of BBN \cite{KKKM-1,KKKM-2,gss,other}.
If the mass gap is sufficiently large, the decay of the sneutrino produces 
high-energy, non-thermal neutrinos. 
Through scattering processes with the background particles, such as 
$\nu_i + \bar{\nu}_{j, {\rm BG}} \to
(e^\pm , \mu^\pm, \tau^\pm)$, $\nu_i + \bar{\nu}_{i, {\rm BG}} \to \pi^+ + \pi^-$ and 
$\nu_i + e_{i, {\rm BG}}^\pm \to \pi^0 + \pi^\pm$, the energetic neutrinos
transfer some parts of their energies to charged particles. The final-state
particles may then photodissociate
or hadrodissociate the elements already produced by standard BBN processes. In 
the case of the charged pion, it can alter the neutron-to-proton ratio if it 
occurs at the beginning of BBN. 
There can also be energy transfer through elastic scattering with electrons and
positrons: $\nu_i + e^\pm \to \nu_i + e^\pm $, and the high-energy $e^\pm$ 
might then initiate
electromagnetic showers. However, at the epoch of interest (when their energies
are ${\cal O}(1)$~MeV) the electron and positron number densities are already low. Therefore
these processes can be neglected.

Subdominant three- and four-body sneutrino decay channels can also be
important, even though their branching ratios are relatively small. This is
because these decays produce charged and/or strongly-interacting particles
directly. These effects had been studied in~\cite{KKKM-1}, where it was
found that the effects of the three- and four-body decays are negligible if their
collective branching ratio is less than about $10^{-6}$.
To estimate this branching ratio, 
we calculate the decay rate for the following process 
\beq
\snu \to \widetilde{G} + \nu + \gamma ,
\eeq
which occurs through neutralinos exchange, with the photon produced via
the photino content of the neutralinos. 
This provides an estimate of the total multi-body decay rate that
should be accurate to within an order of magnitude.
The detailed calculation can be found in the
Appendix. 

\begin{figure}
\begin{center}
\mbox{\epsfig{
file=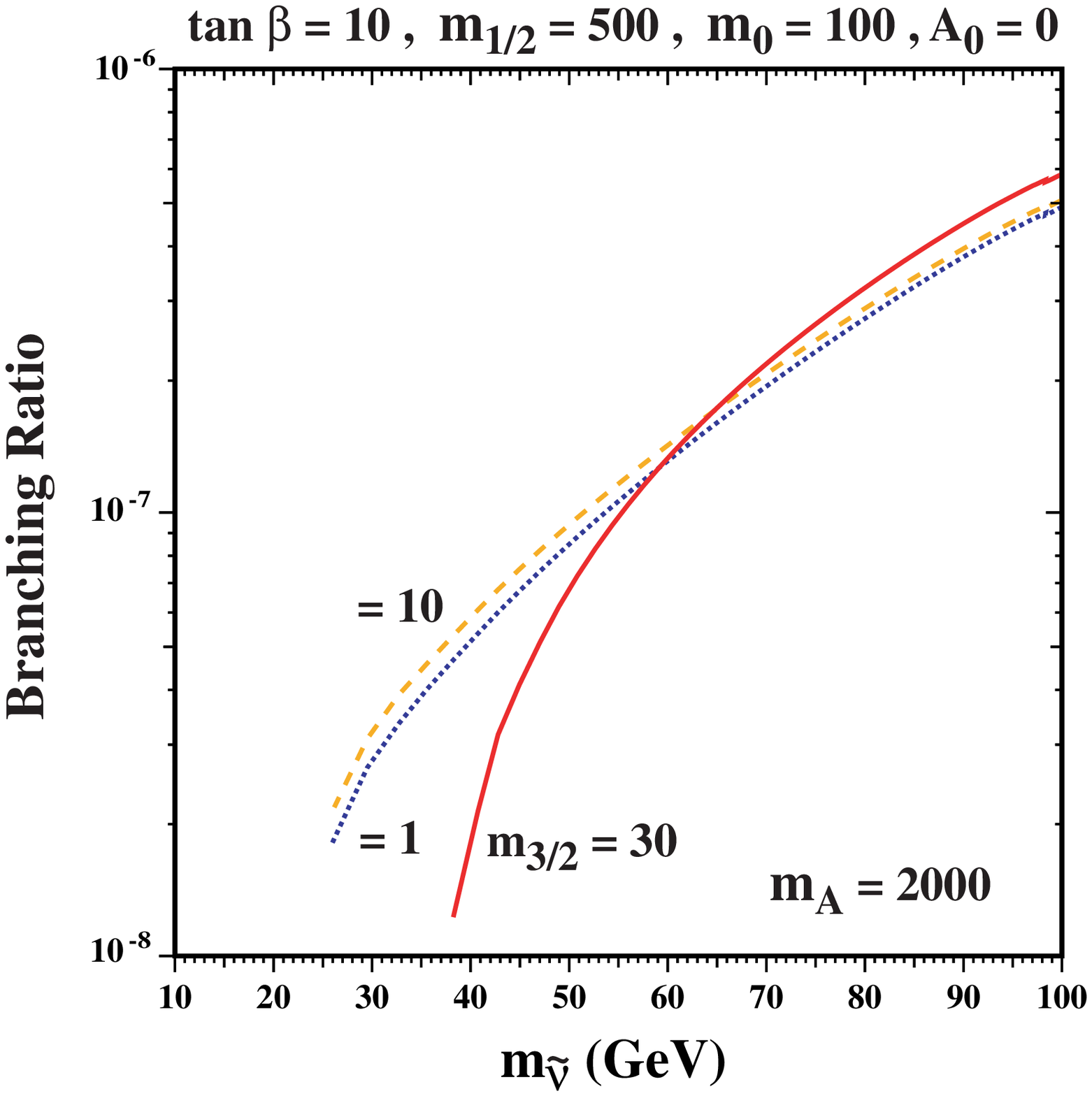,height=8cm}}
\mbox{\epsfig{
file=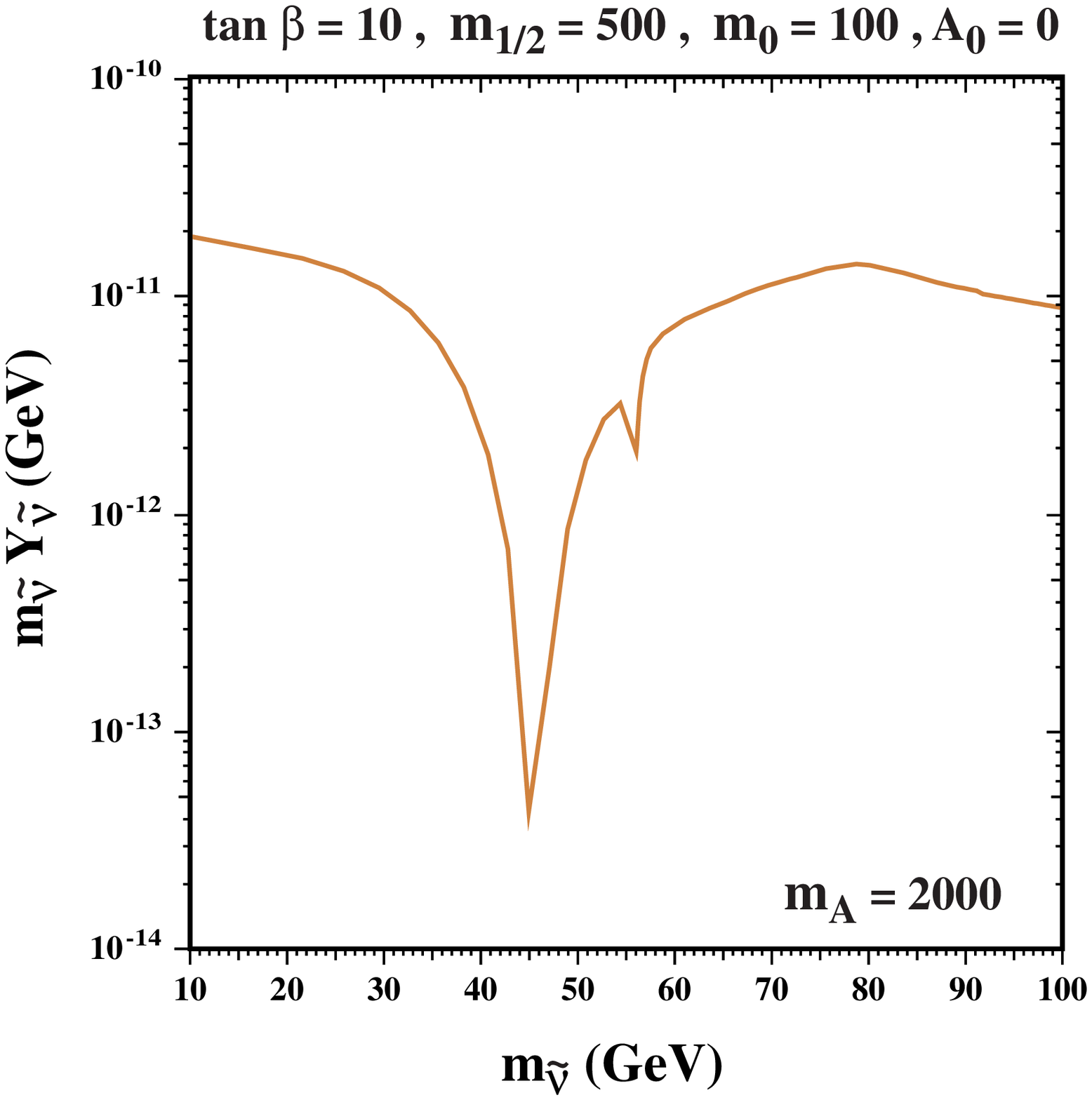,height=8cm}}
\end{center}
\begin{center}
\mbox{\epsfig{
file=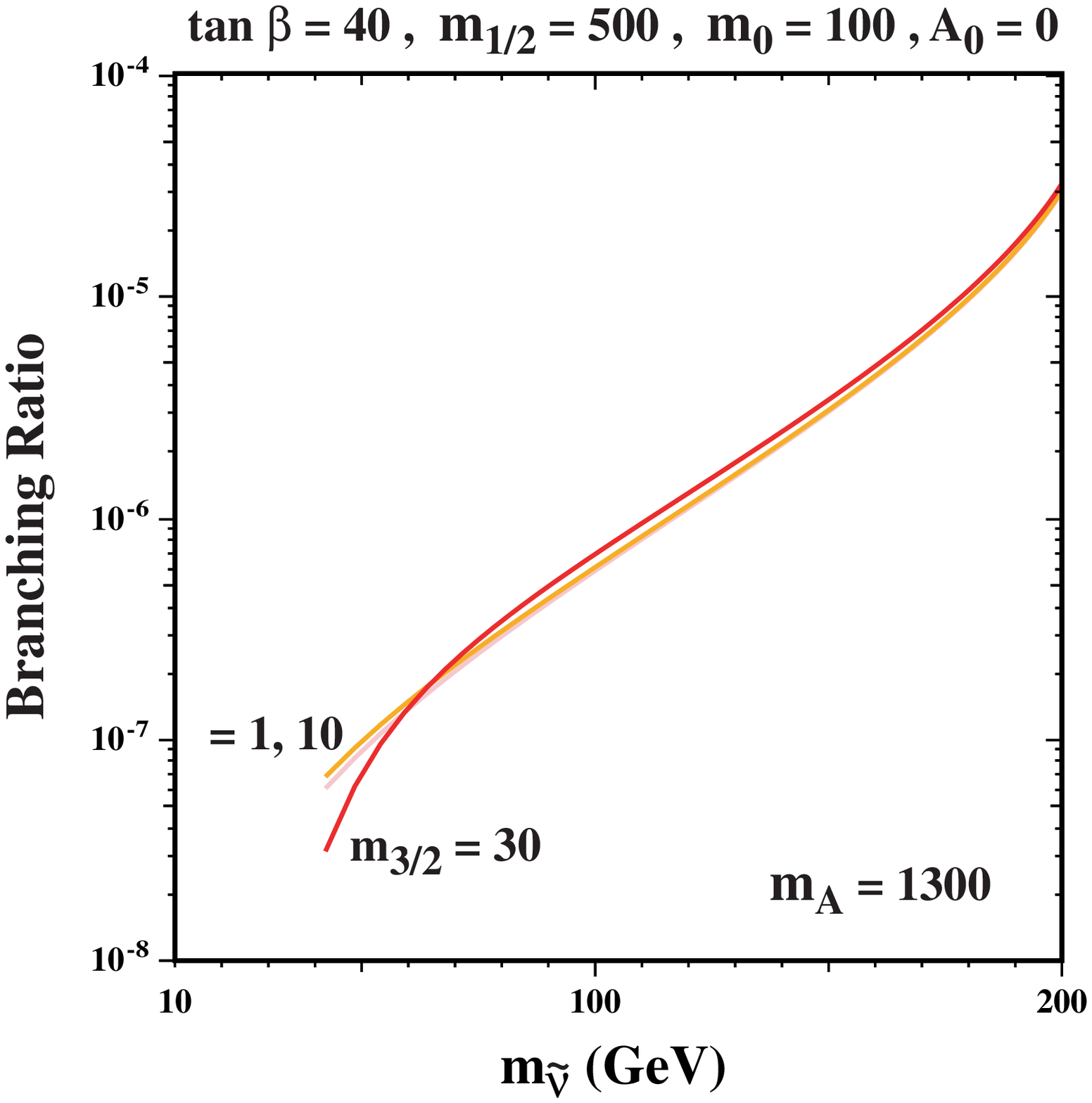,height=8cm}}
\mbox{\epsfig{
file=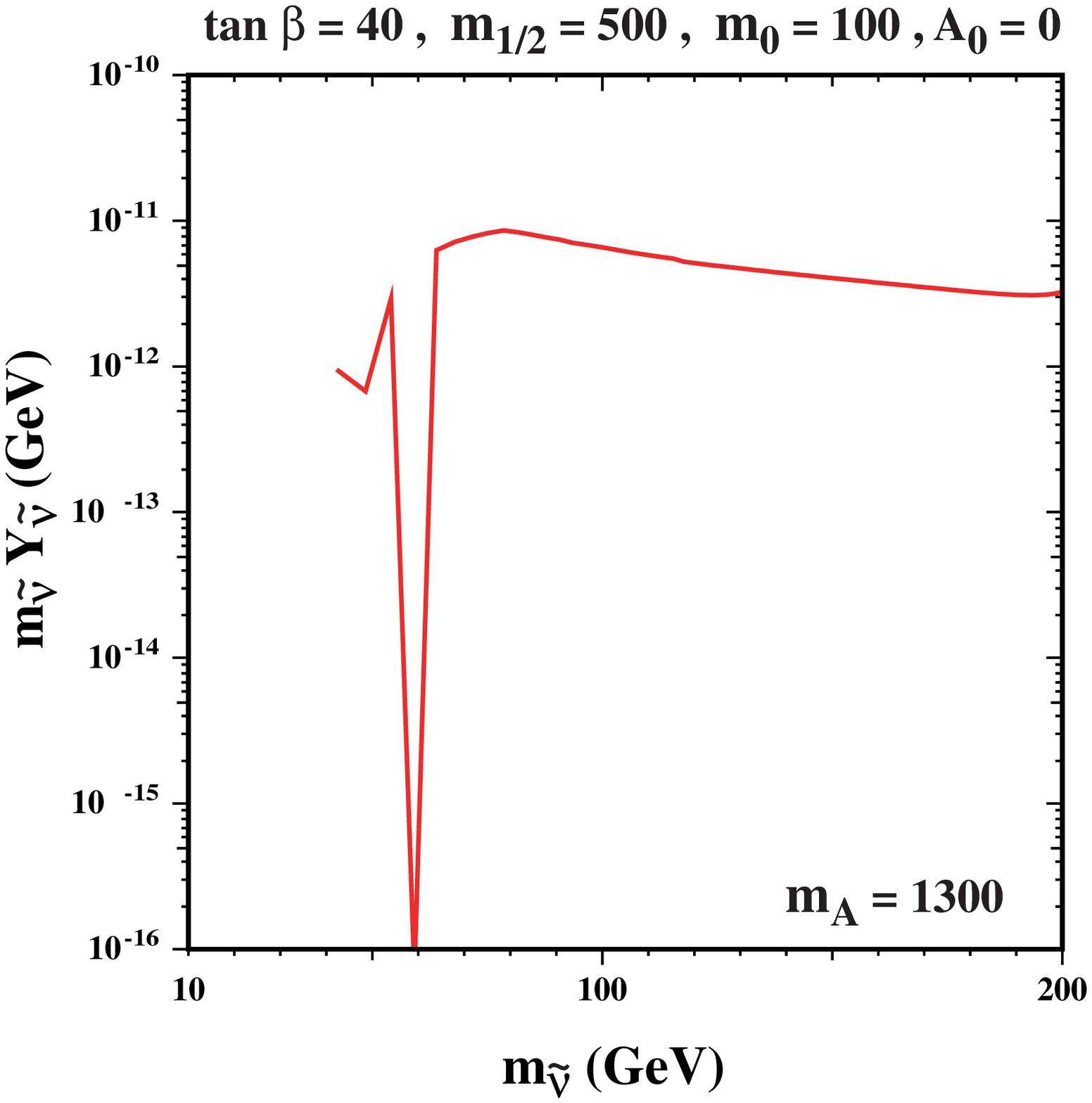,height=8cm}}
\end{center}
\caption{\label{fig:snuBR} \it 
Exploration of the BBN constraints on sample tau-sneutrino NLSP points
with $\mu > 0$: (top)
$m_{1/2} = 500$~GeV, $m_0 = 100$~GeV, $\tan \beta = 10$, $A_0 = 0$
and $m_A = 2000$~GeV [cf, Fig.~\ref{fig:massvsmu}(d)] and (bottom)
$m_{1/2} = 500$~GeV, $m_0 = 100$~GeV, $\tan \beta = 40$, $A_0 = 0$
and $m_A = 1300$~GeV [cf, Fig.~\ref{fig:mumA}(d)]. 
Panels (a, c) display the three-body-decay branching ratios, 
and panels (b, d) the $\tilde{\nu}_{(\tau,e)}$ relic density.} 
\end{figure}

To gauge the possible impact of the BBN constraint, we examine the tau sneutrino
NLSP region in Fig.~\ref{fig:massvsmu}(d), where the sneutrino mass varies
from the LEP lower limit of about 40~GeV up to about 100~GeV. We first consider
the case $m_{1/2} = 500$~GeV, $m_0 = 100$~GeV, $\tan \beta = 10$, $A_0 = 0$
and $m_A = 2000$~GeV shown in the top two panels of Fig.~\ref{fig:snuBR}.
Panel (a) shows the three-body-decay branching ratio for various 
values of $\mgrav = 1, 10, 30$~GeV, corresponding to $\snu$ lifetimes
$\gappeq 10^5$~s. We see that the branching ratio is always very small, 
falling below $10^{-6}$ throughout the range of parameter space considered.
This is consistent with the results of \cite{KKKM-2} which also finds a small
hadronic fraction when the sneutrino mass is $\lappeq 100 $ GeV.
Thus, according to the analysis of~\cite{KKKM-1}, the three-body $\snu$ decay
is too small to affect significantly the successful results of BBN.

Although the hadronic branching ratio is expected to be small, BBN nevertheless sets
a limit on the density of sneutrinos at the time of decay.  
For a 100 GeV sneutrino, all lifetimes are safe so long as 
the quantity
\beq
Y_{\tilde{\nu}} M_{\tilde{\nu}} = \Omega_{\tilde{\nu}} h^2 \times (3.65 \times 10^{-9} \; {\rm GeV}) 
\eeq 
is less than $\mathcal{O}(10^{-11})$ GeV for $B_h = 10^{-3}$ and less than 
$\mathcal{O}(10^{-8})$ for $B_h = 10^{-6}$
where $Y_{\tilde{\nu}}$ is the ratio of the number density of sneutrinos to entropy, $n_{\tilde{\nu}}/s$.
In panel (b) of Fig.~\ref{fig:snuBR}, we display the sneutrino relic
density following freeze-out but prior to decay as $Y_{\tilde{\nu}}
M_{\tilde{\nu}}$.
We see that  $Y_{\tilde{\nu}} M_{\tilde{\nu}}$ is always below about $10^{-11}$ GeV,
with a large dip at $\msnu \sim 45$~GeV due to the $Z$ resonance in 
sneutrino-pair annihilation (with a smaller dip at $\msnu \sim 60$~GeV due to the 
$h$ resonance).  Thus, the sneutrino density is far below the range
where BBN constraints become important for the range of the three-body branching ratio
shown in panel (a) of Fig.~\ref{fig:snuBR}.

The lower panels of Fig.~\ref{fig:snuBR} display the three-body branching ratio and
$Y_{\tilde{\nu}} M_{\tilde{\nu}}$ for another case:
$m_{1/2} = 500$~GeV, $m_0 = 100$~GeV, $\tan \beta = 40$, $A_0 = 0$
and $m_A = 1300$~GeV. We see in panel (c) that the three-body branching ratio is
smaller than $10^{-6}$ for $M_{\tilde{\nu}} < 110$~GeV, and always $< 3 \times 10^{-5}$.
Panel (d) shows that $Y_{\tilde{\nu}} M_{\tilde{\nu}}$ is, again, always below about $10^{-11}$ GeV.
In this case, the prominent dip due to the light $h$ resonance in 
sneutrino-pair annihilation, and the direct-channel $Z$ resonance is less important.
This difference from the previous case is due to the larger value of $\tan \beta$.

These examples are indicative that the sneutrino LSP regions in the NUHM parameter space
are generally safe from BBN constraints. We next check other possible constraints on decaying
sneutrinos.

When the high-energy decay neutrinos are thermalized, their energy is transferred and converted
to radiation. If the sneutrinos decay after about $z = 10^7$ (corresponding to 
a lifetime of $1.8 \times 10^7$ s), then the
photons produced might not have a chance to thermalize, and could show up as
distortion of the CMB black-body radiation spectrum.
These constraints were also considered in \cite{KKKM-1} where it was found
that for lifetimes between $10^7$ and $10^{13}$s, the upper limit on 
$Y_{\tilde{\nu}} M_{\tilde{\nu}}$ is roughly $10^{-9}$ ($10^{-7}$) GeV for $B_h = 10^{-3} (10^{-6})$.
Thus in the parameter space we are interested in, this too is never a serious constraint.

When the sneutrino and gravitino mass are nearly degenerate, the sneutrino
lifetime might be very long ($\gappeq 10^{13}$ s as seen in Fig.~\ref{fig:2body}).
If the sneutrinos have decayed after the time of recombination,
they will produce a diffuse neutrino and photon background. 
In principle, there is then a very strong constraint from water Cerenkov detectors
placing an upper limit on $Y_{\tilde{\nu}} M_{\tilde{\nu}}$ of order $10^{-12} - 10^{-15}$ GeV \cite{gss}.
However, these detectors lose sensitivity for neutrino energies below
about 10 GeV~\cite{Halzen:2002pg}.
Thus we again do not expect a severe constraint placed on the parameter space of interest.

\section{Signatures of Metastable Sneutrinos at Colliders}

We have seen in the previous sections that a sneutrino NLSP
is a generic possibility in the NUHM. It would be metastable, so that its
decays would not be seen at colliders, but the late decays of
relic sneutrinos are not excluded by the available cosmological
constraints. The mass of such a sneutrino NLSP might be as low as
the LEP lower limit~\cite{efos,rpp}.
As a non-decaying
neutral particle, the sneutrino would have a missing-energy signature at colliders.
Distinguishing the sneutrino from other possible origins of such events would
require a search for the heavier states that decay into the sneutrino inside the detector. 

Covi and Kraml~\cite{CK} have studied several scenarios with a sneutrino NLSP
assuming the following mass hierarchies. (a) $\mstau > \mchi > m_{\tilde{\nu}_\tau}$:
in this case, neutralino decays into a neutrino and sneutrino are invisible, and the signatures 
of such decay chains resemble those in a conventional neutralino LSP scenario. However, 
if $\mchi >  m_{\tilde{\nu}_\tau} + m_\tau$, the neutralino can decay into 
$\chi \to \tau \tilde{\nu}_\tau f \bar{f}^\prime$, where the  $f \bar{f}^\prime$ pair is
soft if the mass gap is small.
(b) $\mchi > \mstau > m_{\tilde{\nu}_\tau}$: in this case, besides the invisible 
$\nu + \snu$ decay mode, the neutralino can undergo cascade
decays that might be detectable, such as $\tau + \stau$. 
(c) $\mchi > m_{\tilde{e}_L} > \mstau > m_{\tilde{\nu}_\tau}$: in this case, there are
additional decay channels with neutralino decay into an electron and a selectron which
subsequently decays into leptons and a tau-sneutrino. 
Heavier sparticles might decay via the lightest neutralino, but they might 
also decay directly into the $\snu$, e.g., via $\chi_2 \to \nu + \snu$, or via sleptons,
e.g., $\chi_2 \to \tau + \stau, \stau \to \snu + f + {\bar f^\prime}$.

We see from Fig.~\ref{fig:massvsmu} that there are several possible scenarios for the
sparticle mass spectrum in the NUHM, which are distinct from the standard CMSSM spectrum
as we now enumerate. 

\begin{enumerate}
\item 
$\mchi > m_{\tilde{e}_L} > m_{\tilde{\nu}_e} > \mstau > m_{\tilde{\nu}_\tau}$: \\

This is the mass hierarchy seen in panel (d) of Fig.~\ref{fig:massvsmu} for larger $m_A$. 
The neutralino can decay into
\bear
\chi & \to & \tilde{e}_L + e \nnl 
&&  \tilde{\nu}_e + \nu_e \nnl
&&  \tilde{\tau}_1 + \tau \nnl
&&  \tilde{\nu}_\tau + \nu_\tau
\eear
which, for the first mode, would be followed by 
\bear
\tilde{e}_L & \to & \tilde{\nu}_e + \bar{f}^\prime + f \nnl 
&&  \tilde{\tau}_1 + e + \tau \nnl
&&  \tilde{\tau}_1 + \nu_e + \nu_\tau \nnl
&&  \tilde{\nu}_\tau + e + \nu_\tau \nnl
&&  \tilde{\nu}_\tau + \nu_e + \tau ,
\eear
the second mode by
\bear
\tilde{\nu}_e & \to &  \tilde{\tau}_1 + \nu_e + \tau \nnl
&&  \tilde{\tau}_1 + e + \nu_\tau \nnl
&&  \tilde{\nu}_\tau + \nu_e + \nu_\tau \nnl
&&  \tilde{\nu}_\tau + e + \tau ,
\eear
and the third mode by
\beq
\tilde{\tau}_1 \to  \tilde{\nu}_\tau + \bar{f}^\prime + f
\eeq
In the case of the two-body neutralino decay to stau and tau, the decay rate is
\bear
\Gamma (\chi \to \tilde{\tau}_1 + \tau) &=& \frac{\sqrt{m_\chi^4 + m_\tau^4 + \mstau^4 - 2 (m_\tau^2
\mstau^2 + m_\chi^2 m_\tau^2 + m_\chi^2 \mstau^2) }}{32 \pi m_\chi^3} \nnl &&
\times  
\left( (|C_R|^2 +
|C_L|^2 ) (m_\chi^2 + m_\tau^2 - \mstau^2) - 2 (C_L C_R^\ast + C_R C_L^\ast)
m_\tau m_\chi \right)  \nnl ,&&
\eear
where $C_L$, $C_R$ are the left and right couplings in the neutralino-tau-stau
vertex. There are similar expressions are for the other two-body decay modes. 

\item $\mchi > m_{\tilde{\nu}_\tau} > \mstau > m_{\tilde{e}_L} >
m_{\tilde{\nu}_e}$: \\
This hierarchy occurs for more intermediate values of $m_A$ when
$|\mu|$ is large as seen in panels (b) and (c) of Fig.~\ref{fig:massvsmu}~\footnote{The
viability of such models would require some action to conform with the GUT constraint,
e.g., by constraining inflationary cosmology.}.
The neutralino 2-body decay modes would be the same as in the previous case,
although with different branching ratios. However, the cascade decays are in general
different. In this case, we would have
\bear
\tilde{\nu}_\tau & \to & \tilde{\tau}_1 + \bar{f}^\prime + f \nnl 
&&  \tilde{e}_L + e + \nu_\tau \nnl
&&  \tilde{e}_L + \nu_e + \tau \nnl
&&  \tilde{\nu}_e + \nu_e + \nu_\tau \nnl
&&  \tilde{\nu}_\tau + e + \tau
\eear
\bear
\tilde{\tau} & \to &  \tilde{e}_L + e + \tau \nnl
&&  \tilde{e}_L + \nu_e + \nu_\tau \nnl
&&  \tilde{\nu}_e + \nu_e + \tau \nnl
&&  \tilde{\nu}_e + e + \nu_\tau
\eear
and
\beq
\tilde{e}_L \to  \tilde{\nu}_e + \bar{f}^\prime + f .
\eeq

\item For large $\tan \beta$, e.g. $\tan \beta = 40$
as shown in panel d of Fig.~\ref{fig:mumA}, and large $|\mu|$, the 
third-generation sleptons get larger masses through the Yukawa couplings. Thus we have 
$\mchi >  m_{\tilde{e}_L} > m_{\tilde{\nu}_e}$ for the lightest sparticles.  
In this case, the neutralino cascade decays become simpler,
\bear
\chi & \to & \tilde{e}_L + e \nnl 
&&  \tilde{\nu}_e + \nu_e 
\eear
which, for the first mode, would be followed by 
\beq
\tilde{e}_L  \to  \tilde{\nu}_e + \bar{f}^\prime + f ,
\eeq
whilst the second mode is invisible.

\item There are also other possibilities for narrower region of parameter space,
near where the masses cross each other in Fig.~\ref{fig:massvsmu}:
(a) $ m_{\tilde{\nu}_\tau} > \mstau > m_{\tilde{e}_L} >
\mchi > m_{\tilde{\nu}_e}$; 
(b) 
$ m_{\tilde{\nu}_\tau} > \mstau > 
\mchi > m_{\tilde{e}_L} > m_{\tilde{\nu}_e}$; and 
(c) 
$ m_{\tilde{\nu}_\tau} >  
\mchi > \mstau > m_{\tilde{e}_L} > m_{\tilde{\nu}_e}$.
\end{enumerate}

We extract from these examples a few generic features. As in the cases of many
other scenarios beyond the Standard Model, particularly within the general
framework of supersymmetry, the most prominent signature of a sneutrino NLSP scenario is
likely to be missing energy. However, there would in general be accompanying signatures
that would enable a sneutrino NLSP scenario to be distinguished from other
possibilities. Specifically, one expects to see also events with missing energy
accompanied by leptons. The precise nature of this supplementary signature would,
however, depend on the nature of the sneutrino: ${\tilde \nu_\tau}, {\tilde \nu_\mu}$
or ${\tilde \nu_e}$, and on the hierarchy of heavier sparticle masses. 

In particular,
the relative multiplicities of different charged leptons would depend on the flavour
of the invisible $\snu$, and hence be a useful tool for identifying it. Concretely, in
cases where the parent sparticle has no lepton flavour, as would normally be the case at
the LHC, each ${\tilde \nu_\tau}$ NLSP would be accompanied by an
unmatched $\tau$ or $\nu_\tau$, and each ${\tilde \nu_{e,\mu}}$ NLSP would be 
accompanied by an unmatched electron, $\mu$ or corresponding neutrino. In general,
there would be additional lepton-antilepton pairs with matched flavours.

\section{Summary}

We have analyzed in this paper the possibility of a sneutrino NLSP in NUHM
models with a gravitino LSP. This possibility does not exist in the CMSSM,
but is quite generic in the NUHM, as we have illustrated with various specific
examples. The sneutrino might well be the ${\tilde \nu_\tau}$, but the 
${\tilde \nu_\mu}$ and ${\tilde \nu_e}$ are also possible candidates for the
NLSP. A sneutrino NLSP would be metastable and subject to cosmological
constraints on late-decaying particles, but we have shown that these are not
difficult to respect. There are various different possible scenarios for the
spectrum of sparticles heavier than the sneutrino, which would have distinctive
signatures at colliders. In addition to events  with missing energy carried
away by the invisible $\snu$, there would also be events with accompanying
charged leptons. The flavours of such leptons would help identify the flavour
of the sneutrino NLSP.

As particle physics embarks on the study of the TeV scale with the LHC,
much unknown physics will surely be revealed. Supersymmetry is
occasionally regarded as a `known unknown' in the sense that, whereas we
do not know whether it exists,we think we know what it would look like
if it does exist. This paper reminds us that supersymmetry should rather be
regarded as an `unknown unknown', in the sense that not only do we not
know whether it exists, but we also do not know what it would look like.
In the conventional `known unknown' scenario, the LSP is the lightest
neutralino and supersymmetry would produce missing-energy events.
The latter would also be the signature of a scenario with a gravitino LSP
with a neutralino NLSP, at least in gravity-mediated scenarios. However,
once the `Pandora's box' of a gravitino LSP has been opened, many other
NLSP candidates fly out. In addition to the relatively familiar case of the
lighter stau and the more radical case of the lighter stop, there are other
possibilities including the sneutrino NLSP scenarios discussed here.
All of these scenarios have distinctive features, as illustrated here,
so the LHC and subsequent experiments have good prospects for
detecting and distinguishing between the various `unknown unknowns'.
No `unknown unknown' stone should be left unturned in the search for
supersymmetry.

\section*{Acknowledgments}
\noindent 
The work of K.A.O. was supported in part by DOE grant DE--FG02--94ER--40823. 
Y.S. is grateful for the hospitality of CERN and University of Victoria where
part of this work was done, and thanks  Terrance Figy, Gudrun Heinrich, Stefan
Hesselbach, Emerson Luna and Kazunori Kohri for useful discussions.

\section*{Appendix: Sneutrino Three-Body Decay}

We calculate here the radiative sneutrino three-body decay
\beq
\tilde{\nu}(P) \to \nu(p) + \gamma^\sigma(k) + \widetilde{G}^\mu(P_G)
\eeq
that may arise through the photino content of the neutralino, as
illustrated in the diagram below.

\begin{center} 
  \begin{picture}(140,100)(0,0)
    \DashLine(0,75)(60,75){3} \Text(5,67)[]{$\tilde{\nu}$}
    \ArrowLine(60,75)(120,95) \Text(117,87)[]{$\nu$}
    \ArrowLine(60,20)(60,75) \Text(50,50)[]{$\chi^0_i$}
    \Photon(60,20)(120,50){2.5}{6} \Text(117,39)[]{$\gamma$}
    \ArrowLine(120,0)(60,20) 
    \Photon(120,0)(60,20){2.5}{6} \Text(117,12)[]{$\widetilde{G}$}
  \end{picture}
\end{center}
The invariant amplitude for this decay is
\beq
{\cal M} = \frac{-i \, B_i C_i }{4 M_{\rm Pl} (q^2 - m_i^2)} 
\bar{u}(p) P_R (\slash{q} + m_i) \gamma_\mu \left[ \slash{k},
\gamma_\sigma \right] \Psi^\mu(P_G) \epsilon^\sigma (k) ,
\eeq
where $M_{\rm Pl} = 1/\sqrt{8\pi G_N}$ is the Planck mass, 
$m_i \equiv m_{\chi^0_i}$,  
\beq
q \equiv k + P_G
\eeq
and the dimensionless couplings are
\bear
B_i &\equiv & - \frac{g_2}{\sqrt{2}} ( O_{2 i} - \tan \theta_W O_{1i} )\\
C_i &\equiv & O_{1 i} \cos \theta_W + O_{2 i} \sin \theta_W .
\eear
Ignoring the neutrino mass, we get
\bear
\overline{ \left| {\cal M} \right|^2} &=& 
\frac{8}{3} \frac{k \cdot P_G}{M_{\rm Pl}^2} \sum_{i,j} \frac{C_j^\ast C_i B_j^\ast
B_i}{(q^2 - m_i^2)(q^2 - m_j^2)}  \nonumber \\
&=& \left\{ \frac{k \cdot P_G}{m_G^2} \left[ 2 (P_G \cdot q) (p \cdot q) + (P_G
\cdot p) (m_i m_j - q^2) \right] \right. \nonumber \\
&& \left. \quad \quad + 2 (k \cdot q) (p \cdot q) + (k \cdot p) (m_i m_j - q^2)
\right\}  .
\eear
Use the Dalitz parametrization $m_{12} \equiv p_1 + p_2$, $m_{23} \equiv p_2 + p_3$
with $p_1 = P_G$, $p_2 = k$, $p_3 = p$, we get
\bear
P_G \cdot k &=& \half (m_{12}^2 - m_G^2) \\
k \cdot p &=& \half m_{23}^2 \\
P_G \cdot p &=& \half ( M^2 - m_{12}^2 - m_{23}^2 ) \\
k \cdot q &=& \half (m_{12}^2 - m_G^2 ) \\
p \cdot q &=& \half ( M^2 - m_{12}^2) \\
P_G \cdot q &=& \half ( m_{12}^2 + m_G^2 )
\eear
and
\beq
q^2 = m_{12}^2 .
\eeq
Hereafter, we abbreviate our notation by defining $M \equiv m_{\tilde{\nu}}$ and
$m_G \equiv \mgrav$.

The resulting partial decay rate is~\cite{rpp} 
\beq
d\Gamma = \frac{1}{(2\pi)^3} \, \frac{1}{32 M^3} 
\overline{ \left| {\cal M} \right|^2} \, dm_{12}^2 \, dm_{23}^2 .
\eeq
This can be integrated analytically using the following integration boundaries - for $m_{23}^2$:
$0$ and $(m_{12}^2 M^2 + m_{12}^2 m_G^2 - m_{12}^4 -
M^2 m_G^2 )/m_{12}^2$, and for $m_{12}^2$: $m_G^2$ to
$M^2$. 
The result is
\beq
\Gamma = \frac{1}{768 \pi^3 M_{\rm Pl}^2 M^3} \sum_{i,j} \frac{C_j^\ast
C_i B_j^\ast B_i}{2} \left( I_a(i,j) + I_b(i,j) + I_c(i,j) + I_d(i,j) \right) ,
\eeq
where
\bear
I_a(i,j) & \equiv & \frac{m_i^2 + m_j^2}{m_G^2 (m_i m_j)^4} \left[ 
\sum_{a = 1}^7 \frac{\alpha_a(i,j)}{a} (M^{2 a} - m_G^{2 a}) 
+ \alpha_0(i,j) \ln \left[ \frac{M^2}{m_G^2} \right] \right]  , \\
I_b(i,j) &\equiv & \frac{1}{m_G^2 (m_i m_j)^2} \left[ \sum_{a = 1}^6 
\frac{\alpha_{a+1}(i,j)}{a} (M^{2 a} - m_G^{2 a}) 
+ \alpha_1(i,j) \ln \left[ \frac{M^2}{m_G^2} \right] 
\right.  \nonumber \\
&& \hspace{2cm} 
\left. - \alpha_0(i,j) \left( \frac{1}{M^2} - \frac{1}{m_G^2} \right) \right] , \\
I_c(i,j) &\equiv&  \frac{1}{m_G^2 m_i^4 (m_i^2 - m_j^2)} \left[
\sum_{a=1}^7 \frac{\beta_a(i,j)}{a} \left( (M^2 - m_i^2)^a - (m_G^2 - m_i^2)^a
\right) \right. 
\nonumber \\
&& \hspace{2cm} 
\left. + \beta_0(i,j) \ln \left[ \frac{M^2 - m_i^2}{m_G^2 - m_i^2} \right] \right]  ,
\eear
and 
\beq
I_d(i,j) \equiv I_c(j,i) ,
\eeq
where the auxiliary functions $\alpha$ and $\beta$ are defined below.  
Note that there is no actual singularity when $i=j$, because $I_c + I_d$ is of
the form
\beq
\left( \frac{f(a,b)}{a^2} - \frac{f(b,a)}{b^2} \right) \, \frac{1}{a-b}
\eeq
and in this case, with $m_i = m_j$, we get 
\bear
I_c + I_d &=& \frac{1}{m_G^2 m_i^2}
\nonumber \\
&&  
\left[ - 2 \frac{u^7}{7} + (-13 m_i^2 - 2
\alpha_6 ) \frac{u^6}{6} + ( -35 m_i^4 - 11 m_i^2 \alpha_6 - 2 \alpha_5)
\frac{u^5}{5} \right.
\nonumber \\
&&  
+ ( -49 m_i^6 - 24 m_i^4 \alpha_6 - 9 m_i^2 \alpha_5  - 2 \alpha_4)
\frac{u^4}{4} 
\nonumber \\
&& 
+ ( -35 m_i^8 - 25 m_i^6 \alpha_6 - 15 m_i^4 \alpha_5 
 - 7 m_i^2
\alpha_4 - 2 \alpha_3 ) \frac{u^3}{3} 
\nonumber \\
&&  
+ (-7 m_i^{10} - 10 m_i^8 \alpha_6 - 10
m_i^6 \alpha_5 - 8 m_i^4 \alpha_4 - 5 m_i^2 \alpha_3 - 2 \alpha_2 )
\frac{u^2}{2} \nonumber \\
&&  
+ ( 7 m_i^{12} + 3 m_i^{10} \alpha_6 - 2 m_i^6 \alpha_4 - 3 m_i^4
\alpha_3 - 3 m_i^2 \alpha_2 - 2 \alpha_1 ) u 
\nonumber \\
&&  
+ ( 5 m_i^{14} + 4 m_i^{12} \alpha_6 + 3 m_i^{10} \alpha_5 + 2 m_i^8 \alpha_4 +
m_i^6 \alpha_3 - m_i^2 \alpha_1 - 2 \alpha_0 ) \ln u 
\nonumber \\
&&  \left.
- (m_i^{16} + m_i^{14} \alpha_6 + m_i^{12} \alpha_5 + m_i^{10} \alpha_4 + m_i^8
\alpha_3 + m_i^6 \alpha_2 + m_i^4 \alpha_1 + m_i^2 \alpha_0 ) \frac{1}{u}
\right]_{u = m_G^2 - m_i^2}^{M^2 - m_i^2}    .
\eear
The auxiliary functions are
\bear
\alpha_0(i,j) &\equiv& - 3 m_i m_j m_G^8 M^4 \\
\alpha_1(i,j) &\equiv&  m_i m_j (8 m_G^6 M^4 + 6 M^2 m_G^8) - 3 m_G^8 M^4 \\
\alpha_2(i,j) &\equiv&  8 m_G^6 M^4 + 6 M^2 m_G^8 - m_i m_j (3 m_G^8 + 16 M^2
m_G^6 + 6 M^4 m_G^4 ) \\
\alpha_3(i,j) &\equiv&  m_i m_j (8 m_G^6 + 12 M^2 m_G^4) - 3 m_G^8 - 16 M^2
m_G^6 - 6 M^4 m_G^4 \\
\alpha_4(i,j) &\equiv& m_i m_j ( M^4 - 6 m_G^4) + 8 m_G^6 + 12 M^2 m_G^4  \\
\alpha_5(i,j) &\equiv&  M^4 - 6 m_G^4 - 2 m_i m_j M^2 \\
\alpha_6(i,j) &\equiv&  m_i m_j - 2 M^2 \\
\alpha_7(i,j) &\equiv& 1 ,
\eear
and 
\bear
\beta_0(i,j) &\equiv& \sum_{a=0}^7 m_i^{2a} \alpha_a(i,j)  \\
\beta_1(i,j) &\equiv& \sum_{a=1}^7 a m_i^{2(a-1)} \alpha_a(i,j) \\
\beta_2(i,j) &\equiv& 21 m_i^{10} + 15 m_i^8 \alpha_6 + 10 m_i^6 \alpha_5 + 6
m_i^4 \alpha^4 + 3 m_i^2 \alpha_3 + \alpha_2 \\
\beta_3(i,j) &\equiv&  35 m_i^8 + 20 m_i^6 \alpha_6 + 10 m_i^4 \alpha_5 + 4
m_i^2 \alpha_4 + \alpha_3 \\
\beta_4(i,j) &\equiv&  35 m_i^6 + 15 m_i^4 \alpha_6 + 5 m_i^2 \alpha_5 +
\alpha_4 \\
\beta_5(i,j) &\equiv&  21 m_i^4 + 6 m_i^2 \alpha_6 + \alpha_5 \\
\beta_6(i,j) &\equiv&  7 m_i^2 + \alpha_6 \\
\beta_7(i,j) &\equiv& 1 .
\eear

\end{document}